\documentclass[doublecol]{epl2}
\usepackage{amsmath,amssymb,bm,mathtools}
\usepackage{graphicx}
\usepackage{microtype}
\usepackage[colorlinks=true,linkcolor=blue,citecolor=blue,urlcolor=blue]{hyperref}
\usepackage{xcolor}
\newcommand{\blue}[1]{\textcolor{black}{#1}}

\title{Tuning the strength of emergent correlations in a Brownian gas via batch resetting}

\shorttitle{Emergent correlations via batch resetting}

\author{Gabriele de Mauro\inst{1} \and Satya N. Majumdar\inst{1} \and Gr\'egory Schehr\inst{2}}

\shortauthor{G. de Mauro \etal}

\institute{
  \inst{1} LPTMS, CNRS, Universit\'e Paris-Saclay, 91405 Orsay, France\\
  \inst{2} Sorbonne Universit\'e, Laboratoire de Physique Th\'eorique et Hautes Energies, CNRS UMR 7589, 4 Place Jussieu, 75252 Paris Cedex 05, France
}

\abstract{
We study a gas of $N$ diffusing particles on the line subject to \emph{batch resetting}: at rate $r$, a uniformly random subset of $m$ particles is reset to the origin. Despite the absence of interactions, the dynamics generates a nonequilibrium stationary state (NESS) with long-range correlations. Since standard renewal methods fail for $m<N$, we introduce a new general Fokker--Planck framework with an exact recursive closure of marginal densities. We obtain exact results, both for the NESS and for the time dependence of the correlations, which are valid for arbitrary $m$ and $N$. By varying $m$, the system interpolates between an uncorrelated regime ($m=1$) and the fully synchronous resetting case ($m=N$). For all $1<m<N$, correlations exhibit a non-monotonic time dependence due to the emergence of an intrinsic decorrelation mechanism. In the stationary state, the correlation strength can be tuned by varying $m$, and it displays a transition at a critical value $N_c=6$. \blue{Our predictions extend to any spatial dimension $d$, where the critical value $N_c=6$ remains unchanged, and they are experimentally testable in existing optical-trap setups.}}

\makeatletter
\let\epl@pacsmissing\@empty
\let\epl@pacsset\@empty
\makeatother

\begin{document}

\maketitle

Stochastic resetting was introduced in \cite{EM2011,EM2011b} based on a simple idea: the dynamics of a system is interrupted at random times and restarted from a fixed initial condition. This mechanism is very general and it can be applied to any deterministic or stochastic dynamics. It has attracted substantial theoretical and experimental interest because of its nontrivial effects on the behavior of various systems. For example, by breaking detailed balance, the stochastic resetting drives the system to a nonequilibrium stationary state (NESS) \cite{EM2011,EM2011b,EMS2020,PKR2022,GuptaJ2022,NagarG2023,GMS2014,BKP2019,MMS2020,CZ2024,AN2026}. Also, stochastic resetting can dramatically enhance the efficiency of search processes \cite{EM2011,EM2011b,EMS2020,PKR2022,EM2014,KMSS2014,R2016,PKE2016,PR2017,CS2018,P2018,EM2019,BS2020,PKR2020,SB2021,DMS2022,BMS2023,DeBruyneM2023,GPPM2023,MFC24,PPPPSL2024,TKRR2025,HM2025,MoriMahad2025,EvansRay2025,DKMS2026,BMSP2025_ts}. 
More recently, striking dynamical effects have been uncovered in systems where many degrees of freedom (DOF) are all reset simultaneously. In particular, a one-dimensional gas of $N$ independent Brownian particles, all reset simultaneously to the origin at a rate $r$, develops strong long-range correlations and reaches a NESS that, though strongly correlated, remains analytically tractable for physically measurable observables~\cite{Biroli2023}. These correlations arise purely from the dynamics: although the particles never interact, the \emph{simultaneity} of the resets couples all DOF and induces effective correlations. Such dynamically emergent correlations (DEC) have been observed in several classical \cite{Biroli2023,BLMS2024,BKMS2024,SabM2024,BMSP2025_ts,BMS2025,MMS2025,BoyerM2026,BMS2026_fpg,DeMauro2025,Galla2026,Olsen2026,MS2026_DEC,Bressloff2026} and quantum \cite{MCPL2022,KMS2025,SLP2025} systems, in which the DOF do not interact directly, but become strongly correlated through a shared fluctuating environment. For a recent perspective on DEC see \cite{MS2026_DEC}.

In the examples mentioned above, all DOF are coupled to the same single fluctuating environment that acts globally. This naturally raises the question of what survives of this phenomenology when environmental fluctuations act only on a subset of the DOF at the same time. We show below that this change induces a qualitatively different physics in the system and its analytical treatment becomes highly nontrivial. Previous works have considered partitioned systems, in which the DOF are split into two subsets: one is repeatedly reset, whereas the other is never reset.
It was shown in \cite{AMG2025,MCG2024,Majumder2026} that one can control the stationary state of the full system by a repeated resetting of only one of the subsystems.
However, this requires the two subsystems to have an extensive number of long-range interactions. What we study here is fundamentally different: in our system, direct interactions between the DOF are completely absent. In addition, all DOF are coupled to the same fluctuating environment, but each fluctuation acts only on a randomly chosen subset rather than on the whole system. Over long times, these repeated fluctuations ensure that every DOF is intermittently affected.
\begin{figure}[t]
    \centering
    \includegraphics[width=1\columnwidth]{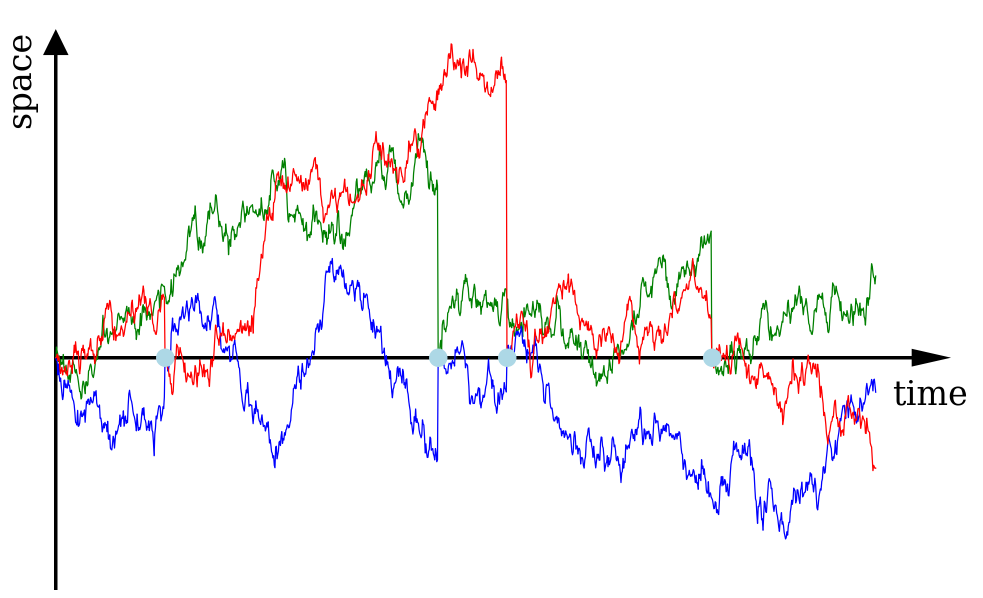}
    \caption{Schematic representation of batch resetting for $N=3$ and $m=2$. Four resetting events are shown (light-blue dots).}
    \label{fig:proces}
\end{figure}

\blue{Our model should also not be confused with the so-called partial resetting protocols~\cite{TFR2022,BFHMS2024}, mostly studied for single-particle processes, in which a DOF is reset only partially, e.g. $x\to ax$ with $|a|<1$, rather than exactly to the origin.}

\blue{In this Letter, we introduce {\it batch resetting}, a minimal setting to study this phenomenon in which each resetting event acts only on a randomly selected subset of DOF. It can be viewed as a simple yet nontrivial model for a broad class of systems in which an external fluctuating drive or environment acts on different parts of the system at different fluctuation events.}
We consider a one-dimensional system of $N$ diffusing particles (with diffusion constant $D$) that undergo stochastic resetting at a constant rate $r$. At each reset event, only $m$ particles, chosen uniformly at random, are reset to the origin, while the remaining $N-m$ particles keep diffusing (see Fig.~\ref{fig:proces}). Thus, the fluctuating environment is effectively generated by the stochastic resetting of a randomly selected subset of $m$ particles, where the subset is independently re-sampled at every event.

It is important to stress that the problem becomes mathematically far more difficult as soon as $m<N$. When all $N$ particles are reset together ($m = N$), the dynamics obeys a simple renewal equation that directly yields the stationary state \cite{EMS2020,Biroli2023}. For any intermediate value $1 \leq m < N$ , however, this renewal structure is lost: since only a subset of particles is reset during a reset event, the system never returns to a fully renewed configuration. As a result, the renewal equation breaks down and the standard 
methods used for simultaneous resetting ($m=N$) no longer apply. 

\blue{However, for any $1 \leq m \leq N$, we uncover that the batch-resetting dynamics has a special solvable structure: the exact Fokker--Planck equation for the $n$-point marginal PDF involves only $k$-point marginals with $k \leq n$. This is in sharp contrast with generic many-body systems, where the evolution of low-order marginals typically generates a BBGKY-type hierarchy~\cite{Kirkwood}, in which lower-order marginals are coupled to higher-order ones, making the problem unsolvable without a closure approximation. 
Here, instead, no closure approximation is needed: starting from the one-point marginal, all higher-order marginals can be obtained recursively and exactly, at all times.
This exact recursive closure is the key conceptual and technical advance of this paper, allowing us to compute the time-dependent two-point correlation function for arbitrary parameters. As shown later in the Supplemental Material (SM)~\cite{SM}, the same property holds for diffusion in arbitrary spatial dimension, and we expect it to be a generic and robust property of systems with batch resetting, extending beyond simple diffusive dynamics.}

\blue{Using this framework, we show that batch resetting generates robust long-range attractive correlations even for $m<N$. Here, long-range should be understood in the sense of global, all-to-all correlations between particles, rather than correlations characterized by a slow spatial decay: by permutation symmetry, all distinct pairs are correlated in the same way, and there is no characteristic spatial length scale over which these correlations decay. Importantly, their strength can be tuned through the single parameter $m$, as the system interpolates between an uncorrelated stationary state at $m=1$ and the fully synchronous limit at $m=N$.} Moreover, new phenomena appear for $m<N$. First, the system develops non-monotonic correlations in time due to a new decorrelation mechanism with a characteristic timescale that diverges as $m\to N$. Its origin is discussed below.
\begin{figure*}[ht]
    \centering
    \includegraphics[width=\textwidth]{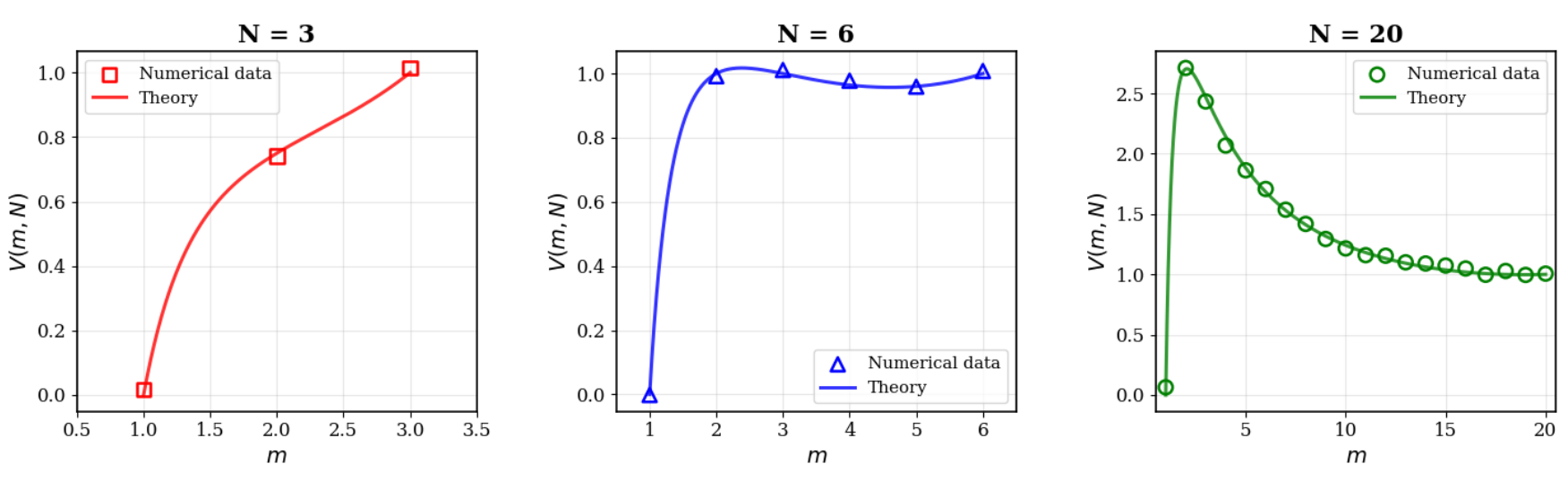}
    \caption{
    Plot of the function $V(m,N)$ (Eq.~\eqref{eq:C2stat}) compared with numerical simulations, shown as a function of $m$ for different values of $N$.
    For $N<N_c=6$, $V(m,N)$ increases monotonically with $m$, whereas for $N>N_c$
    it becomes non–monotonic: the global maximum shifts from $m=N$ to $m=2$ and a shallow local minimum develops at $m=N-1$. This minimum is clearly visible in the central panel but it is also present for all $N\geq N_c$.
    At the critical value $N=N_c$, the points $m=2,3,N$ all give the same maximal value $V(m,N_c)=1$. The rightmost point corresponds to fully synchronous resetting~\cite{Biroli2023} for all values of $N$.}
    \label{fig:V(m,N)}
\end{figure*}
\blue{Second, these strong, attractive correlations persist all the way to the NESS, where their unnormalized stationary amplitude displays a transition as a function of $m$ when the total number of particles crosses the critical value $N_c=6$. This transition results from the competition between single-particle fluctuations and two-particle correlations.}

Recent experiments have made stochastic resetting directly accessible in the lab. Colloidal particles diffusing in water can now be reset using laser-generated optical traps, both individually \cite{Besga2020,Tal-Friedman2020,Faisant2021,GinotBechinger2026} and in groups \cite{Vatash2025,biroli2025exp}. Notably, in Ref. \cite{biroli2025exp}, each particle is confined in its individual optical trap, and the stiffnesses of all the traps are synchronously modulated at random times, effectively implementing simultaneous resetting events. This setup offers a direct route to test our predictions for batch resetting, as each particle being in its own trap allows selective resetting of any subset of $m$ particles.

To study the case $m < N$ we develop a Fokker--Planck approach. We denote by $P_m(\mathbf{x},t)$ the joint probability density function (JPDF) of the system, where $\mathbf{x}=(x_1,...,x_N)$ are the positions of the particles and the subscript $m$ indicates that exactly $m$ particles are reset at each event.  It is convenient to introduce its Fourier transform (FT)
\begin{equation}\label{eq:FP_Kspace}
    \tilde{P}_m(\mathbf{k},t)
    = \int d\mathbf{x}\, e^{i\mathbf{k}\cdot\mathbf{x}}\, P_m(\mathbf{x},t),
    \quad 
    \mathbf{k}=(k_1,\ldots,k_N),
\end{equation}
which we refer to as the FT--JPDF. We also define the $n$-point marginals of the JPDF as $P_m^{n}(x_1,\ldots,x_n,t)=\!\int\!dx_{n+1}\cdots dx_N\,P_m(\mathbf{x},t)$, with FT 
$\tilde P_m^{n}(k_1,\ldots,k_n,t)=\!\int\!dx_1\cdots dx_n\,e^{i(k_1x_1+\cdots+k_nx_n)}P_m^{n}(x_1,\ldots,x_n,t)$. We recall that, in this dynamics, 
at each resetting event, a subset $S_m\subseteq\{1,\ldots,N\}$ of size $|S_m|=m$ is selected uniformly at random. For those
particles whose labels belong to $S_m$, the position is set to zero. The goal is to write a Fokker--Planck equation for this batch resetting protocol. This can be done either in the real space or in the Fourier space (which is slightly simpler). 
In the SM \cite{SM}, we derive the Fokker--Planck equation for both representations. Here, for  
convenience, we describe the form in Fourier space
\begin{align}\label{eq:FP_inkspace}
\partial_t \tilde{P}_m(\mathbf{k},t)
    &= -\big( D k^2 + r \big)\, \tilde{P}_m(\mathbf{k},t) \nonumber \\
    &\quad+ \frac{r}{\binom{N}{m}}
    \sum_{\substack{S_m}}
    \tilde{P}^{N-m}_m\!\big(\mathbf{k}_{S_m^{\mathrm{c}}}, t\big) \,,
\end{align}
while the one in direct space is given in the SM~\cite{SM}.
The first term in \eqref{eq:FP_inkspace}, 
$- D k^2 \tilde{P}_m(\mathbf{k},t)$, corresponds to standard diffusion, where $k^2 = |\mathbf{k}|^2 = \sum_{i=1}^N k_i^2
$. The second contribution, $- r \tilde{P}_m(\mathbf{k},t)$, represents the loss of probability induced by resetting events occurring at rate~$r$.
The last term encodes the gain of probability coming from all possible resetting events. Recalling that the positions of particles whose 
labels belong to $S_m$ are set to zero, it follows that, in Fourier space, one needs to set the corresponding Fourier momentum $k_i=0$. This is a direct consequence of the definition of FT, given in \eqref{eq:FP_Kspace}.
It is convenient to also define the set $S_m^{\mathrm{c}} = \{1,\ldots,N\} \setminus S_m$ and the vector $\mathbf{k}_{S_m^{\mathrm{c}}}$, which is obtained from $\mathbf{k}=(k_1,\ldots,k_N)$ by setting to zero the components $\{k_i\}_{i\in S_m}$. 
Accordingly, $\tilde{P}^{N-m}_m(\mathbf{k}_{S_m^{\mathrm{c}}},t)$ denotes the $(N-m)$-point marginal in Fourier space, evaluated at $\mathbf{k}_{S_m^{\mathrm{c}}}$. 
Averaging this contribution over all
$\binom{N}{m}$ possible reset subsets of size $m$ produces the factor $1/\binom{N}{m}$ in \eqref{eq:FP_inkspace}. The notation used in \eqref{eq:FP_inkspace} may look a bit abstract, but it becomes clearer by explicitly considering a simple example, such as $N=3$ and $m=2$, which we work out in detail in the SM~\cite{SM}.

Equation \eqref{eq:FP_inkspace} has an interesting structure, namely, that the equation for the $n$-point marginal of the FT--JPDF involves only marginals of order up to $n$ on the right hand side, and not higher-order ones. Consequently, the full time-dependent solution of the $n$-point marginal $\tilde P_m^{n}(k_1,\ldots,k_n,t)$ can, in principle, be obtained recursively. 
The same structure also applies to the stationary problem, where the calculations are simpler because one only needs to set the time derivative in \eqref{eq:FP_inkspace} to zero. This leads to a closed set of equations characterizing the stationary FT--JPDF and all its marginals.
In particular, it provides an exact solution for the JPDF when $m=N-1$ and $m=1$ (see SM~\cite{SM}), in addition to the known result for $m=N$.
For a general $1<m<N-1$, an exact solution remains possible in principle, but cumbersome to find in practice.
However, the recursive structure of \eqref{eq:FP_inkspace} already allows us to obtain exactly the marginals of low order that are needed for the correlation functions. We show here how this works for the one- and two-point marginals in the stationary state, namely $\tilde{P}_m^1(k)$ and $\tilde{P}_m^2(k_1,k_2)$. In the SM~\cite{SM}, using the same method, we also derive the full time dependence of these two marginals at all times $t$, i.e. $\tilde{P}_m^1(k,t)$ and $\tilde{P}_m^2(k_1,k_2,t)$. 

Setting $\partial_t \tilde P_m=0$ in \eqref{eq:FP_inkspace} leads to
\begin{equation}\label{eq:stat_tildePk}
    \tilde{P}_m(\mathbf{k})
= \frac{\displaystyle \frac{r}{\binom{N}{m}}
        \sum_{S_m}
        \tilde{P}_m^{N-m}(\mathbf{k}_{S_m^{\mathrm{c}}})}
     {\,r + D \sum_{i=1}^N k_i^2\,}.
\end{equation}
This expression relates the Fourier transform of the NESS for a general $m$ to the Fourier transform of its marginals. Since the equation is valid for any vector $\mathbf{k}$, we can choose to calculate it in $\mathbf{k}=(k,0,...,0)$ in order to obtain a closed equation for $\tilde P_m(\mathbf{k}=(k,0,\dots,0))$, which, given the definition of FT in Eq. \eqref{eq:FP_Kspace}, coincides with the FT of the one-point marginal $\tilde P_m^1(k)$. This closed equation can be easily solved. Repeating the same argument, we can evaluate \eqref{eq:FP_inkspace} in $(k_1,k_2,0,\dots,0)$, which gives an expression involving only $\tilde P_m^1(k)$ and $\tilde P_m^2(k_1,k_2)$. Since we already know 
$\tilde P_m^1(k)$, we can solve for $\tilde P_m^2(k_1,k_2)$ and then invert the Fourier transform. This method leads to the exact expressions
\begin{equation}\label{eq:P1tilde}
\tilde{P}_m^1(k) = \frac{1}{1+\ell_m^2 k^2} \,,\, {\rm where} \quad \ell_m^2 = \frac{D}{\tilde{r}}\,, \quad \tilde{r} = \frac{m}{N}r \;,
\end{equation}
and
\begin{equation}\label{eq:P2tilde}
\tilde{P}_m^2(k_1,k_2) = \frac{\,(2-\beta) + (\beta-1) \left[\tilde{P}_m^1(k_1) + \tilde{P}_m^1(k_2)\right]}{ \,  \, \beta + \ell_m^2(k_1^2 + k_2^2)} \;,
\end{equation}
which are valid for all $1\leq m\leq N$ and for all $N\geq 2$.

The parameter $\beta=\frac{2N-m-1}{N-1}\in [1,2]$ interpolates between $\beta = 1$ as $m\to N$ and $\beta = 2$ as $m\to 1$. The two coefficients $(2-\beta)$ and $(\beta-1)$ appearing in Eq.~\eqref{eq:P2tilde} have a simple probabilistic interpretation,  which will be important below. Fix two particles $i$ and $j$ and condition on particle $i$ being reset during a given event. Since the remaining $m-1$ reset particles are chosen uniformly among the other $N-1$ particles, the probability that particle $j$ is also reset is $2-\beta=\frac{m-1}{N-1}$,
whereas the complementary probability that particle $j$ is not reset is $ \beta-1=\frac{N-m}{N-1}$. Thus, the factor $2-\beta$ can be interpreted as the probability of a simultaneous reset of the pair, while $\beta-1$ is the probability of an asynchronous reset, in which only one of the two particles is reset.

By inverting the FTs in \eqref{eq:P1tilde} and \eqref{eq:P2tilde} we obtain
\begin{eqnarray}\label{eq:1-point_marg}
    P_m^1(x) = \frac{1}{\ell_m} \mathcal{R}\left( \frac{x}{\ell_m} \right), \quad \mathcal{R}(u) = \frac{1}{2}e^{-|u|},
\end{eqnarray}
while $P_m^2(x_1,x_2)$ takes the scaling form
\begin{equation}
    P_m^2(x_1,x_2)=\frac{1}{\ell_m^2}\,
    \mathcal P_\beta\!\left(\frac{x_1}{\ell_m},\frac{x_2}{\ell_m}\right),
\end{equation}
and the explicit expression for $\mathcal P_\beta(u,v)$, involving modified Bessel functions, is rather lengthy and is therefore reported in the SM~\cite{SM}.

Since all particles are statistically equivalent in our system, the average density of particles $\rho_N^m(x) = \frac{1}{N} \langle \sum_{i=1}^{N} \delta(x - x_i) \rangle$ coincides with the one-point marginal $P_m^1(x)$ given above, namely $\rho_N^m(x)=P_m^1(x)=\frac{1}{\ell_m}\mathcal{R}(x/\ell_m)$. This function decays exponentially over the length scale $\ell_m = \sqrt{\frac{ND}{mr}} \sim O(1/\sqrt{m/N})$, which decreases as the fraction of reset particles increases. However, the scaling function $\mathcal{R}(u)$ is independent of both $m$ and $N$, and is therefore identical to the fully synchronous case~\cite{Biroli2023}. In this sense, all one-particle observables are insensitive to whether resets involve all particles or only a subset, up to the rescaling $r \to \tilde r = \frac{m}{N}r$.
In contrast, the two-point marginal $P_m^2(x_1,x_2)$ acquires a qualitatively different structure for $m < N$ (see SM \cite{SM}), leading to new phenomena in the two-particle observables, such as the emergence of nontrivial two-point correlations, as discussed below.

To quantify the DEC between two particles, it is natural to consider the standard two-point correlation function $\langle x_i(t) x_j(t) \rangle- \langle x_i(t) \rangle\,\langle x_j(t) \rangle$. However, due to the $x_i \to - x_i$ symmetry in this system, this standard correlator is identically zero. Hence, to detect the correlations, the simplest and most natural quantity is the next-order correlator $C_{ij}= \langle x_i^2(t) x_j^2(t) \rangle- \langle x_i^2(t) \rangle \,\langle x_j^2(t) \rangle\,, \;\, i,j\in\{1,\ldots,N\},$
where $\langle\cdot\rangle$ denotes the average with respect to the JPDF $P_m(\mathbf{x},t)$. Using the permutation symmetry, it follows that 
$C_{ii} = \mathcal{C}_1(t) \,\, \forall\, i,$ and $C_{ij} = \mathcal{C}_2(t) \,\, \forall \, i \neq j $. Note that $\mathcal{C}_2(t)$ is a measure of correlations, but its magnitude is still sensitive to the overall fluctuation scale of the system. Since $\mathcal{C}_1(t)$ provides the corresponding single-particle fluctuation scale, it is natural to consider the normalized correlator
\begin{eqnarray}\label{eq:a_C2/C1}
    a(t;m,N) = \frac{\mathcal{C}_2(t)}{\mathcal{C}_1(t)}
    = \frac{\langle x_i^2(t) x_j^2(t) \rangle - \langle x_i^2(t) \rangle\,\langle x_j^2(t) \rangle}{\langle x_i^4(t) \rangle - \langle x_i^2(t) \rangle^2} \;.
\end{eqnarray}
One can show that $0\leq a(t;m,N) \leq 1$, interpolating between the completely uncorrelated case ($a=0$) and the fully correlated case ($a=1$) where $x_i^2=x_j^2$ for all pairs $(i,j)$. These quantities were studied previously for the fully synchronous resetting case $m=N$ \cite{Biroli2023,DeMauro2025,BoyerM2026}. We now show that they can be computed exactly also for the batch resetting protocol, for any $1\leq m\leq N$, thus enabling a direct comparison with the fully synchronous case.

There are two equivalent ways to obtain $\mathcal C_1(t)$, $\mathcal C_2(t)$ and $a(t;m,N)$. The first one uses the Fokker--Planck equation in direct space and is illustrated in the SM~\cite{SM}. The second one, which we describe here, uses the time-dependent one- and two-point marginals $\tilde P_m^1(k,t)$ and $\tilde P_m^2(k_1,k_2,t)$. These can be obtained exactly, starting from \eqref{eq:FP_inkspace}, with the same recursive method used above for the stationary marginals. Once these two functions are known, all the required moments follow by differentiation:
\begin{equation}\label{eq:moments_from_FT}
\begin{aligned}
\langle x_i^2(t)\rangle
&=
-\left.\partial_k^2\tilde P_m^1(k,t)\right|_{k=0},
\\
\langle x_i^4(t)\rangle
&=
\left.\partial_k^4\tilde P_m^1(k,t)\right|_{k=0},
\\
\langle x_i^2(t)x_j^2(t)\rangle
&=
\left.
\partial_{k_1}^2\partial_{k_2}^2
\tilde P_m^2(k_1,k_2,t)
\right|_{k_1=k_2=0}.
\end{aligned}
\end{equation}
Defining $\tilde r=(m/N)r$ and $\beta=(2N-m-1)/(N-1)$, this gives
\begin{align}
\mathcal C_1(t)
&=
\frac{4D^2}{\tilde r^2}\,
f_1(\tilde r t),
\label{eq:C1fullexpression}
\\
\mathcal C_2(t)
&=
\frac{4D^2}{\tilde r^2}\,
f_2(\tilde r t;\beta),
\label{eq:C2fullexpression}
\\
a(t;m,N)
&=
A(\tilde r t;\beta),
\label{eq:Afullexpression}
\end{align}
where
\begin{align}
f_1(z)
&=
5-4e^{-z}-6ze^{-z}-e^{-2z},
\label{eq:f1_def}
\\
f_2(z;\beta)
&=
2\left[
\frac{1-e^{-\beta z}}{\beta}
-
\frac{e^{-z}-e^{-\beta z}}{\beta-1}
\right]
-1-e^{-2z}+2e^{-z},
\label{eq:f2_def}
\\
A(z;\beta)
&=
\frac{f_2(z;\beta)}{f_1(z)} .
\label{eq:A_def}
\end{align}
In the fully synchronous limit $m\to N$, i.e. $\beta\to 1$, the function $f_2(z;\beta)$ converges to $f_2(z;1)=1-2ze^{-z}-e^{-2z}$, which recovers the known result for fully synchronous resetting~\cite{DeMauro2025}.

We first consider $\mathcal{C}_2(t)$ and in particular its stationary limit $\mathcal{C}_2^*=\mathcal{C}_2(t\to\infty)$. It follows from \eqref{eq:C2fullexpression} that
\begin{equation}\label{eq:C2stat}
    \mathcal{C}_2^*\!=\!\frac{4D^2}{r^2}V(m,N), \,\,\,V(m,N)=\frac{N^2(m-1)}{m^2(2N-m-1)}. 
\end{equation}
In the fully synchronous case ($m=N$) one recovers the known results $\mathcal{C}_2^*=\frac{4D^2}{r^2}$ \cite{Biroli2023}.
The function $V(m,N)$, shown in Fig.~\ref{fig:V(m,N)}, satisfies $V(N,N)=1$ and $V(1,N)=0$ for all $N$, and it is strictly positive for any $N>1$ and $1<m<N$. 
This shows that the attractive correlations characteristic of fully synchronous resetting persist for all $1<m<N$. Moreover, varying $m$ provides a simple way to tune the strength of the correlations. In addition, we uncover an intriguing transition in the behavior of the stationary correlations as a function of $m$ when the total number of particles crosses $N_c=6$. 
This transition is initially surprising. Since $\mathcal{C}_2^*$ is a connected two-particle correlator, one might naively expect its largest amplitude to be reached in the fully synchronous limit $m=N$, where every reset event involves all particles.
However, as shown in Fig.~\ref{fig:V(m,N)}, for $N<N_c$ the 
correlation strength increases monotonically with $m$, while for $N> N_c$ it becomes non-monotonic and the global maximum shifts from $m=N$ to $m=2$. For $N\geq N_c$, the function also develops a shallow local minimum at $m=N-1$, before increasing again to the fully synchronous value $V(N,N)=1$, as illustrated in the central panel of Fig.~\ref{fig:V(m,N)}.
At $N=N_c$ the maximum is not unique, since at $m=2,3,6$ the function $V(m,N_c)$ attains the same maximal value $1$.
This nontrivial dependence on $m$ arises from the competition between two effects to which the correlator $\mathcal{C}_2^*$ is sensitive: single-particle fluctuations and two-particle correlations. 
In particular, increasing $m$ makes each reset event involve more particles, so correlations are injected into a larger group at once. 
However, a larger value of $m$ also means that each particle is reset more frequently and therefore explores a smaller region between two of its own resets, of size $\ell_m \sim O\!\left(1/\sqrt{m/N}\right)$.
Thus, as $m$ increases, these two trends compete: the contribution to $\mathcal{C}_2^*$ coming from correlations grows, since more particles are simultaneously reset, while the contribution coming from single-particle fluctuations decreases, as the explored region $\ell_m$ becomes smaller. 
Their interplay gives rise to the observed non-monotonic behavior when $N \ge 6$.
This becomes clearer if we normalize the correlations by the typical fluctuations of a single particle, which are captured by the object $\mathcal{C}_1(t)$. This removes the dependence on the exploration range $\ell_m$ and isolates the sole effect of resetting larger groups. In these units, the stationary correlator takes the form $a(t\!\to\!\infty;m,N)=\frac{\mathcal{C}_2(t\to \infty)}{\mathcal{C}_1(t\to \infty)}
= \frac{1}{5}\,\frac{m-1}{\,2N-m-1\,},$ which is strictly monotonic in $m$ for any fixed $N$. 

\begin{figure}[ht]
    \centering
    \includegraphics[width=0.95\columnwidth]{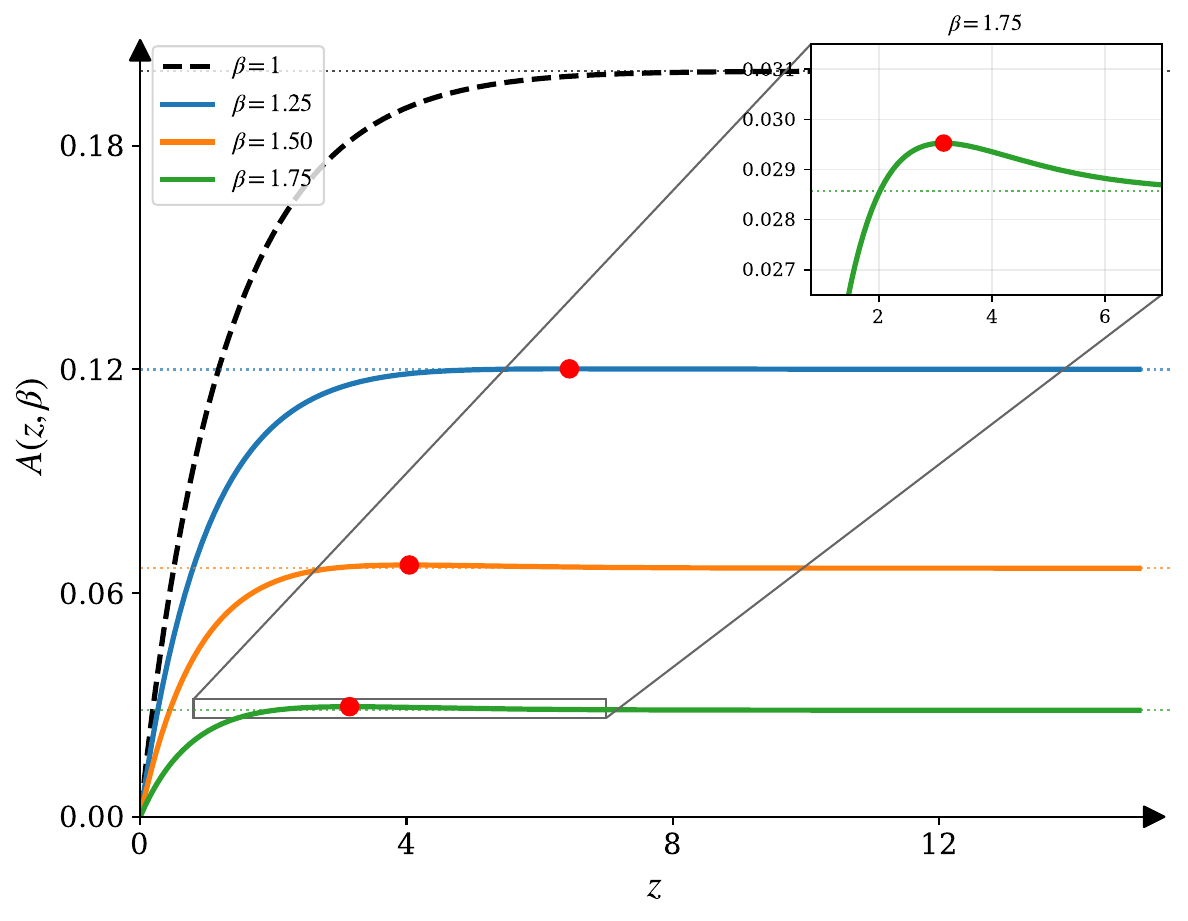}
    \caption{Scaling function $A(z;\beta)$ given in Eq.~\eqref{eq:A_def}, shown for several
    values of $\beta=\frac{2N-m-1}{N-1}\in[1,2]$. 
    For $1<m<N$ ($1<\beta<2$), $A(z;\beta)$ develops a maximum at a finite time $z$
    (indicated by red dots), while for $m=N$ ($\beta=1$) it increases monotonically.
    For any $1<\beta<2$, $A(z;\beta)$ decays exponentially fast to its stationary value
    $(2/\beta-1)/5$. A comparison to numerical simulations, with $\beta = 1.5$, is shown in Fig. S4 in \cite{SM}.}
    \label{fig:Azbeta}
\end{figure}

The second key result of this Letter concerns the time dependence of the function $a(t;m,N)$, given in \eqref{eq:Afullexpression}. It takes the scaling form $a(z=\tilde r t;m,N) \equiv A(z;\beta),$ where $A(z;\beta)$ depends on $m$ and $N$ only through $\beta=(2N-m-1)/(N-1)$. As expected, for $m=1$ ($\beta=2$) the function vanishes identically: the particles are noninteracting, and no pair of particles is ever reset together. In contrast, for $m=N$ ($\beta=1$) it increases monotonically with $z$ (see Fig.~\ref{fig:Azbeta}). A key finding of this work is that this behavior changes qualitatively as soon as $m<N$. For any $1<m<N$ ($1<\beta<2$), the function $A(z;\beta)$ becomes non-monotonic: it first grows (correlation--building phase), then decreases (decorrelation phase), and finally saturates to the stationary value $A(z\to \infty;\beta)= \frac{1}{5}\left(\frac{2}{\beta}-1 \right)$ (see Fig.~\ref{fig:Azbeta}), developing a maximum at a finite time $z=z^*(\beta)$. 

This non-monotonicity originates from an additional decorrelation mechanism which is absent for $m=N$: when $m<N$, one particle may be reset while the other is not, thereby erasing their mutual correlation. This mechanism introduces an additional timescale, which diverges as $\beta\to1^+$. Its origin can be made explicit by deriving a closed evolution equation for $A(z;\beta)$. As shown in the SM~\cite{SM}, one obtains
\begin{equation}\label{eq:deffeqA}
    \frac{dA(z;\beta)}{dz}
    =
    -\Big[(\beta-1)+\omega(z)+\phi(z)\Big]A(z;\beta)
    +(2-\beta)\phi(z),
\end{equation}
with $\omega(z)=4(1-e^{-z})/f_1(z)$, and $\phi(z)=(1-e^{-z})^2/f_1(z)$,
where $f_1(z)$ is defined in Eq. \eqref{eq:f1_def}.
Solving Eq.~\eqref{eq:deffeqA} with $A(0;\beta)=0$ gives back the explicit expression of $A(z;\beta)$ in Eq.~\eqref{eq:A_def}, and hence the curves shown in Fig.~\ref{fig:Azbeta}. More importantly, Eq.~\eqref{eq:deffeqA} makes the physical origin of the maximum transparent. The negative terms proportional to $A(z;\beta)$ are loss terms: they erase pre-existing correlations. The term $-\omega(z)A$ comes from diffusion, since between reset events the two particles diffuse independently and their correlations are progressively washed out. The term $-\phi(z)A$ comes from resetting itself: whenever a particle is reset, its previous trajectory is forgotten, and this loss of memory tends to erase the correlations that had been built before.
However, resetting can also create correlations when the two particles are reset together. In this case, the two particles are brought back to the same position, and correlations are reinjected into the system. This reinjection is encoded in the source term $(2-\beta)\phi(z)$. Indeed, $2-\beta=(m-1)/(N-1)$ is the conditional probability that, given that one particle is reset, the other one is also reset in the same event.
Finally, the term $-(\beta-1)A$ is an additional decorrelation channel which is present only when $m<N$. It corresponds to asynchronous reset events, in which one of the two particles is reset while the other one is not. In this case, the reset erases correlations without reinjecting them. Indeed, $\beta-1=(N-m)/(N-1)$ is the conditional probability of such an asynchronous reset event. This extra loss term vanishes in the fully synchronous limit $\beta\to1$, explaining why the non-monotonic behavior disappears for $m=N$. Thus, the maximum of $A(z;\beta)$ results from the competition between correlation injection by simultaneous reset events and correlation loss due to diffusion, memory erasure by resetting, and asynchronous reset events.
A more detailed analysis of Eq.~\eqref{eq:deffeqA} is given in the SM~\cite{SM}.

\begin{figure}[t]
    \centering
    \includegraphics[width=0.95\columnwidth]{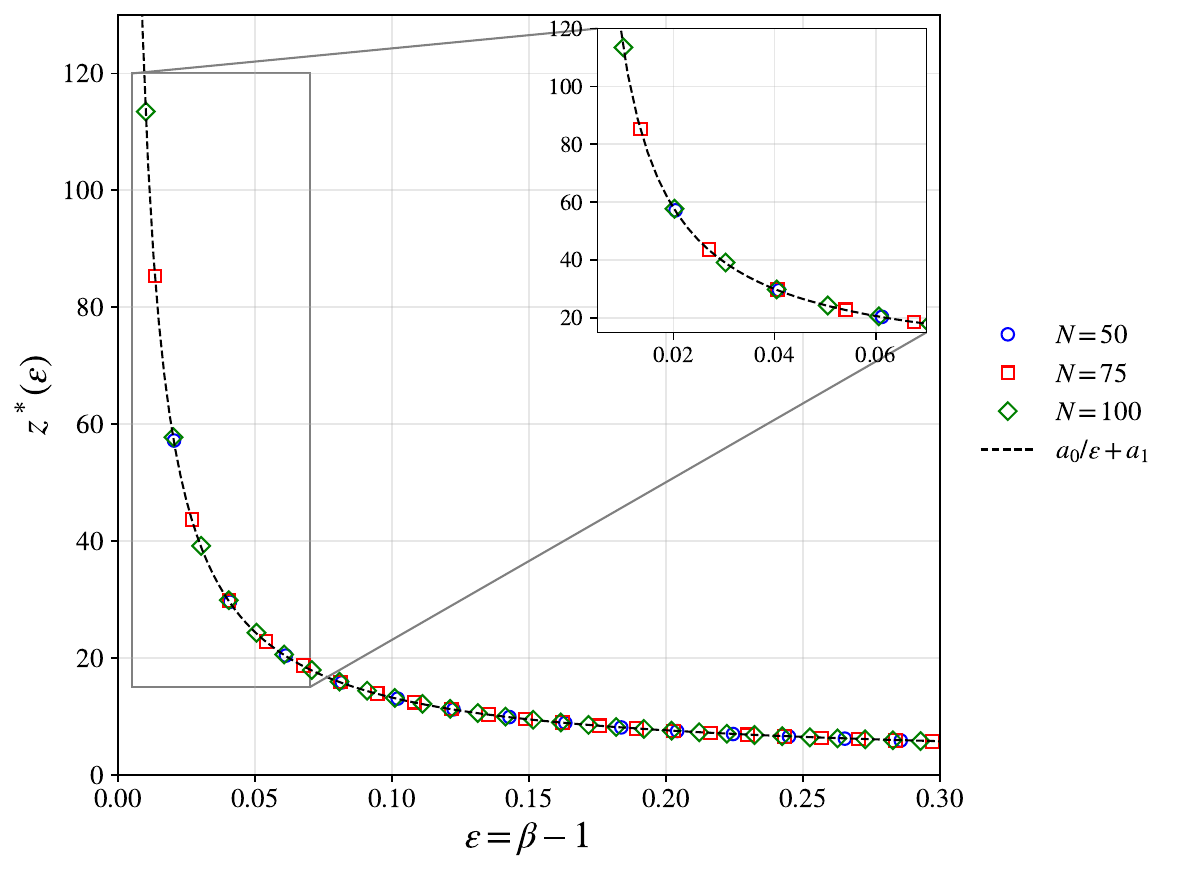} 
    \caption{Position $z^*(\beta=1+\epsilon)$ of the maximum of $A(z;\beta)$ as a function of $\epsilon=\beta-1=\frac{N-m}{N-1}$. The dots show numerical determinations obtained from the exact expression in Eq.~\eqref{eq:A_def}, while the dashed line shows the small-$\epsilon$ asymptotic prediction given in Eq.~\eqref{eq:zstar_eps_app}. The inset zooms into the asymptotic regime $\epsilon\ll1$, where this expansion is derived. The main panel shows that the same asymptotic expression remains a good approximation well beyond the strictly small-$\epsilon$ regime.}
    \label{fig:maxA_numerics}
\end{figure}

We now turn to the position $z^*(\beta)$ of the maximum. Although a closed expression for $z^*(\beta)$ is hard to find for general $\beta$, analytic progress can be made in the limits $\beta\to1^+$ and $\beta\to2^-$.
Close to the fully synchronous limit $m=N$ (i.e. $\beta=1$), we write $\epsilon=\beta-1=\frac{N-m}{N-1}$. In the large-$N$ limit, $\epsilon$ may be treated as a continuous small parameter. From Fig.~\ref{fig:Azbeta} and from the physical argument discussed above, we expect the position of the maximum to diverge as $\epsilon\to0$. Using the ansatz $z^*(\epsilon)=a_0/\epsilon+a_1+O(\epsilon)$ in the extremum condition $\partial_z A(z;\beta)=0$ gives $a_0=1.1262\ldots$ and $a_1=2$, hence
\begin{equation}
\label{eq:zstar_eps_app}
    z^*(\epsilon)\simeq \frac{1.1262\ldots}{\epsilon}+2.
\end{equation}
The detailed derivation of Eq.~\eqref{eq:zstar_eps_app} is given in the SM~\cite{SM}. Figure~\ref{fig:maxA_numerics} shows that this asymptotic prediction agrees well with the numerical determination of the maximum over a broad range of values of $\epsilon$. Equation~\eqref{eq:zstar_eps_app} is consistent with the physical interpretation given above: $\epsilon=\beta-1$ is the conditional probability that, given that one particle is reset, the other one is not. Thus, the timescale associated with this asynchronous decorrelation mechanism diverges as $1/\epsilon$ in the fully synchronous limit.

In the opposite limit $\beta\to2^-$, we set $\beta=2-\delta$, with $\delta=\frac{m-1}{N-1}$. Expanding $A(z;2-\delta)$ to first order in $\delta$ and imposing the extremum condition gives a finite limiting position of the maximum,
\begin{equation}
\label{eq:zstar_beta2_app}
    z^*(\beta)\to \bar z=2.6495\ldots,
    \qquad \beta\to2^- .
\end{equation}
The detailed derivation of Eq.~\eqref{eq:zstar_beta2_app} is also given in the SM~\cite{SM}. Thus, although the amplitude of $A(z;\beta)$ vanishes as $\beta\to2^-$, the position of its maximum remains finite.

Very recently, a similar non-monotonic behavior of $A(z;\beta)$ was found in a system where $N$ diffusing particles are reset simultaneously to previously visited positions, thus introducing an explicit temporal memory kernel \cite{BoyerM2026}. There, the non-monotonicity appears only below a certain threshold of the control parameter and was characterized mostly through numerical analysis. Our batch resetting protocol provides a complementary example: non-monotonic temporal correlations arise for all $1<m<N$ (i.e., $1<\beta<2$) without the need for an explicit memory kernel, and the position $z^*(\beta)$ of the maximum can be investigated analytically in the limits $\beta\to 1^+$ and $\beta \to 2^-$.

For finite $N$, the discussion above sheds light on the correlations between individual particles ensuing from the batch resetting protocol. It is natural to wonder what happens in the large $N$ limit, where our system resembles a strongly correlated gas of particles on a line. For such a continuum many-body system, it is natural to coarse-grain the system and define a local empirical density $\hat \rho(x) = \frac{1}{N}\sum_{i=1}^N \delta(x-x_i)$. 
The natural way to probe the correlations in this coarse-grained system is to study the density-density correlation function or its Fourier transform, known as the spectral form factor. We show in the SM \cite{SM} that these observables can be computed exactly in our system and they exhibit interesting new behaviors that can also be traced back to the batch resetting protocol. 
 
To conclude, batch resetting generates strong and tunable correlations in a diffusive gas. One-particle observables remain identical to the fully synchronous case, up to a rescaling of the resetting rate. In contrast, two-particle observables display qualitatively new behaviors, including non-monotonic temporal correlations and a transition in the stationary correlation strength at $N_c=6$ as a function of $m$. Our exact solution provides a unified framework to interpret these effects and yields quantitative predictions. 

\blue{These predictions may be directly testable in the optical-trap setup of Ref.~\cite{biroli2025exp}, which can implement batch resetting, in principle, by modulating only $m$ out of $N$ traps (with up to $N=8$). This geometry would avoid the problem of particle collisions during the return phase even for $m<N$, since different particles remain confined around distinct trap centers. Residual hydrodynamic interactions may remain, but they are expected to be subdominant when the typical excursion between resets, $\ell_m$, is small compared with the trap spacing~\cite{biroli2025exp}. Finite return-time corrections should also be negligible if the return stage is sufficiently short, although incorporating them explicitly in the theory remains a natural next step. See Ref.~\cite{EMS2020} for a discussion of finite return-time in standard stochastic resetting.}

\blue{Additionally, our analysis extends to higher spatial dimensions. All computations can be carried out exactly in any dimension $d$, as detailed in Sec.~S10 of the SM~\cite{SM}. For any $d$, the normalized correlation function $A(z;\beta)$ remains non-monotonic in $z$, while the function $V(m,N)$ is unchanged and therefore displays the same transition at $N_c=6$. This demonstrates the robustness of our results.}

\blue{From a theoretical perspective, it would be interesting to extend this approach to random batch sizes and to study the effect of batch resetting on multi-agent search processes.}

\acknowledgments
We acknowledge support from ANR Grant No. ANR-23-CE30-0020-01 EDIPS.

\end{document}


\title{Supplemental Material for\\
Tuning the strength of emergent correlations in a Brownian gas via batch resetting}

\author{Gabriele de Mauro}
\affiliation{LPTMS, CNRS, Universit\'e Paris-Saclay, 91405 Orsay, France}

\author{Satya N. Majumdar}
\affiliation{LPTMS, CNRS, Universit\'e Paris-Saclay, 91405 Orsay, France}

\author{Gr\'egory Schehr}
\affiliation{Sorbonne Universit\'e, Laboratoire de Physique Th\'eorique et Hautes Energies, CNRS UMR 7589, 4 Place Jussieu, 75252 Paris Cedex 05, France}

\maketitle
\setcounter{tocdepth}{2} 
\tableofcontents

\setcounter{secnumdepth}{2}
\renewcommand{\thesection}{S\arabic{section}}
\renewcommand{\thesubsection}{S\arabic{section}.\arabic{subsection}}
\renewcommand{\theequation}{S\arabic{equation}}
\renewcommand{\thefigure}{S\arabic{figure}}
\renewcommand{\thetable}{S\arabic{table}}

\newpage

\section{Illustrative example: $N=3$ and $m=2$}

As a first instructive example, we consider a system of \(N=3\) particles in which \(m=2\) of them are reset at each reset event, with a constant rate $r$. Then, in the next section, we generalize the problem to arbitrary values of $m$ and $N$.
The evolution of the joint probability density (JPDF) \(P_{m=2}(x_1,x_2,x_3,t)\) is governed by the Fokker--Planck equation
\begin{equation}\label{eq:FP_x_N3_m2}
\begin{aligned}
\frac{\partial}{\partial t} P_{m=2}(x_1,x_2,x_3,t)
&= D \sum_{i=1}^3
\frac{\partial^2}{\partial x_i^2}
 P_{m=2}(x_1,x_2,x_3,t)- r\, P_{m=2}(x_1,x_2,x_3,t)\\
&\quad + \frac{r}{3} \Bigg[
\delta(x_1)\,\delta(x_2)
\int_{-\infty}^{\infty}\!\!\dd y_1 \dd y_2\,
P_{m=2}(y_1,y_2,x_3,t)\\
&\qquad\qquad
+ \delta(x_1)\,\delta(x_3)
\int_{-\infty}^{\infty}\!\!\dd y_1 \dd y_3\,
P_{m=2}(y_1,x_2,y_3,t)\\
&\qquad\qquad
+ \delta(x_2)\,\delta(x_3)
\int_{-\infty}^{\infty}\!\!\dd y_2 \dd y_3\,
P_{m=2}(x_1,y_2,y_3,t)\Bigg].
\end{aligned}
\end{equation}
The first term on the right-hand side of \eqref{eq:FP_x_N3_m2} describes the independent diffusion of the three particles with diffusion coefficient $D$.
The second term is a loss term accounting for the fact that, with rate $r$, a reset event occurs and the system leaves the configuration $\mathbf{x}=(x_1,x_2,x_3)$.
The remaining terms in the square brackets represent the gain due to reset events.
At each reset, two particles out of three are chosen uniformly at random, hence each of the three possible pairs is selected with probability $1/3$.
For instance, the contribution
\begin{equation}\label{eq:gfgfgf}
    \frac{r}{3}\,\delta(x_1)\,\delta(x_2)
\int_{-\infty}^{\infty}\!\!\dd y_1 \dd y_2\,
P_{m=2}(y_1,y_2,x_3,t)
\end{equation}
corresponds to events where particles $1$ and $2$ are reset to the origin.
The Dirac delta functions $\delta(x_1)\delta(x_2)$ enforce that they are brought at $x_1=x_2=0$, while the integral over $y_1,y_2$ sums over all possible pre-reset positions of these two particles, keeping particle $3$ fixed at $x_3$.
The other two terms in the brackets are interpreted in the same way, with the roles of the three particles permuted. 
It is convenient to take a Fourier transform (FT) of Eq.~\eqref{eq:FP_x_N3_m2}, defining
\begin{equation}\label{eq:defofFT3part}
\tilde{P}_{m=2}(k_1,k_2,k_3,t)
=\int \!\! \int \dd\mathbf{x}\, e^{i\mathbf{k}\cdot\mathbf{x}}\, P_{m=2}(x_1,x_2,x_3,t),
\qquad \mathbf{k}=(k_1,k_2,k_3),
\end{equation}
which gives
\begin{equation}\label{eq:FP_k_N3_m2}
\begin{aligned}
\frac{\partial}{\partial t}\,\tilde P_{m=2}(k_1,k_2,k_3,t)
&= -D\,(k_1^2+k_2^2+k_3^2)\,
    \tilde P_{m=2}(k_1,k_2,k_3,t)- r\,\tilde P_{m=2}(k_1,k_2,k_3,t)\\
&\quad + \frac{r}{3}\Big[
\tilde P_{m=2}(0,0,k_3,t)
+ \tilde P_{m=2}(0,k_2,0,t)
+ \tilde P_{m=2}(k_1,0,0,t)
\Big].
\end{aligned}
\end{equation}
We now introduce the one-point marginal
\begin{equation}\label{eq:def1pointmarginal}
P^1_{m=2}(x,t)
= \int_{-\infty}^{\infty}\!\!\dd x_2 \int_{-\infty}^{\infty}\!\!\dd x_3\,
P_{m=2}(x,x_2,x_3,t),
\end{equation}
and its FT
\begin{equation}\label{eq:def1pointmargfourierspace}
\tilde P^1_{m=2}(k_1,t)
= \int_{-\infty}^{\infty}\!\!\dd x_1 \, e^{ik_1 x_1} \, P^1_{m=2}(x_1,t)=\tilde P_{m=2}(k_1,0,0,t).
\end{equation}
The last equality in \eqref{eq:def1pointmargfourierspace} follows directly from the definition of the FT in \eqref{eq:defofFT3part}. Indeed, setting a component \(k_i = 0\) in \eqref{eq:defofFT3part} is equivalent to marginalizing the JPDF \(P_m(x_1,x_2,x_3)\) over the corresponding spatial variable \(x_i\). As a result, once the FT of the full JPDF, \(\tilde P_{m=2}(k_1,k_2,k_3,t)\), is known, the FT of any marginal distribution can be obtained simply by setting the appropriate components \(k_i\) to zero.
Since all particles play the same role in the dynamics, we also have \(\tilde P^1_{m=2}(k,t)=\tilde P_{m=2}(0,k,0,t)=\tilde P_{m=2}(0,0,k,t)\). This observation will be repeatedly used in the following sections to derive closed equations for the marginal distributions.
Using the definitions in \eqref{eq:def1pointmarginal} and \eqref{eq:def1pointmargfourierspace} we can rewrite~\eqref{eq:FP_x_N3_m2} and \eqref{eq:FP_k_N3_m2} respectively as
\begin{equation}
\begin{aligned}
\frac{\partial}{\partial t} P_{m=2}(x_1,x_2,x_3,t)
&= D \sum_{i=1}^3
\frac{\partial^2}{\partial x_i^2}
 P_{m=2}(x_1,x_2,x_3,t)- r\, P_{m=2}(x_1,x_2,x_3,t)\\
&\quad + \frac{r}{3} \Bigg[
\delta(x_1)\,\delta(x_2)
P^1_{m=2}(x_3,t)
+ \delta(x_1)\,\delta(x_3)
P^1_{m=2}(x_2,t)
+ \delta(x_2)\,\delta(x_3)
P^1_{m=2}(x_1,t)\Bigg],
\end{aligned}
\end{equation}
and
\begin{equation}\label{eq:FP_k_N3_m2_BIS}
\begin{aligned}
\frac{\partial}{\partial t}\,\tilde P_{m=2}(k_1,k_2,k_3,t)
= -\left(r+D|k|^2\right)\,
    \tilde P_{m=2}(k_1,k_2,k_3,t)+ \frac{r}{3}\Big[
\tilde P^1_{m=2}(k_3,t)
+ \tilde P^1_{m=2}(k_2,t)
+ \tilde P^1_{m=2}(k_1,t)
\Big],
\end{aligned}
\end{equation}
where $|k|^2=k_1^2+k_2^2+k_3^2$.
By setting the time derivative in \eqref{eq:FP_k_N3_m2_BIS} to zero, we can obtain an equation for the stationary JPDF in Fourier space, namely
\begin{equation}\label{eq:N3_statstate_kspace}
    \tilde P_{m=2}(k_1,k_2,k_3)
    =
    \frac{\frac{r}{3}\left[ \tilde{P}_{m=2}^1(k_1)+ \tilde{P}_{m=2}^1(k_2)+ \tilde{P}_{m=2}^1(k_3) \right]}
    {r+D(k_1^2+k_2^2+k_3^2)}.
\end{equation}
Using the observation made above, namely that setting components of $\mathbf{k}$ to zero in the Fourier-transformed JPDF yields the FT of the corresponding marginals, we can now evaluate \eqref{eq:N3_statstate_kspace} at $\mathbf{k}=(k,0,0)$. This gives a closed equation for the one-point marginal
\begin{equation}\label{eq:1pmarg_N3}
 \tilde P^1_{m=2}(k)
 =\frac{\frac{r}{3}\left[2+ \tilde P^1_{m=2}(k)\right]}{D\,k^2+r}.
\end{equation}
Solving for \(\tilde P^1_{m=2}(k)\) yields
\begin{equation}\label{eq:N3_P1tilde}
    \tilde P^1_{m=2}(k)=\frac{1}{1+\ell_2^2 \, k^2},
    \qquad
    \ell_2=\sqrt{\frac{3D}{2r}}.
\end{equation}
By inverting the FT, we obtain
\begin{equation}
    P^1_{m=2}(x)
    =\frac{1}{2\pi}\int_{-\infty}^{\infty}\!\!\dd k \, e^{-ikx}\, \tilde P^1_{m=2}(k)
    = \frac{1}{\ell_2}\, \mathcal{R}\!\left(\frac{x}{\ell_2} \right),
    \qquad
    \mathcal{R}(u)=\frac{1}{2}e^{-|u|}.
\end{equation}
We can proceed similarly by evaluating \eqref{eq:N3_statstate_kspace} at \(\mathbf{k}=(k_1,k_2,0)\).
Defining the two-point marginal in Fourier space as \(\tilde P^2_{m=2}(k_1,k_2)\equiv\tilde P_{m=2}(k_1,k_2,0)\), we get
\begin{equation}\label{eq:N3_P2tilde}
    \tilde P^2_{m=2}(k_1,k_2)
    =\frac{\frac{r}{3}\left[ \tilde{P}_{m=2}^1(k_1)+ \tilde{P}_{m=2}^1(k_2)+ 1\right]}
    {r+D(k_1^2+k_2^2)},
\end{equation}
which is a closed equation, since \(\tilde{P}_{m=2}^1(k)\) is already known from \eqref{eq:N3_P1tilde}.
To invert the FT in \eqref{eq:N3_P2tilde}, we use the identity
\begin{equation}
    \frac{1}{(2\pi)^2}\int_{-\infty}^{\infty}\!\!\dd k_1
    \int_{-\infty}^{\infty}\!\!\dd k_2\;
    \frac{e^{-i(k_1 x_1 + k_2 x_2)}}{\,r + D(k_1^2 + k_2^2)\,}
    =\frac{1}{2\pi D}\,
K_0\!\left(
    \sqrt{\frac{r}{D}}\;\sqrt{x_1^{\,2}+x_2^{\,2}}\right),
\end{equation}
where \(K_0(z)\) is the modified Bessel function of the second kind of order \(0\).
This relation leads to
\begin{equation}\label{eq:P2_x_N3_m2_explicit}
\begin{aligned}
P^2_{m=2}(x_1,x_2)
&= \frac{r}{3}\,\frac{1}{2\pi D}\,
K_0\!\left(\sqrt{\frac{r}{D}}\sqrt{x_1^2+x_2^2}\right)\\
&\quad + \frac{r}{3}\frac{1}{2\pi D}
\int_{-\infty}^{\infty}
\frac{\dd y}{2\ell_2}\,e^{-|y|/\ell_2}\, K_0\!\left(\sqrt{\frac{r}{D}}\sqrt{(x_1-y)^2+x_2^2}\right)\\
&\quad+  \frac{r}{3}\frac{1}{2\pi D} \int_{-\infty}^{\infty}
\frac{\dd y}{2\ell_2}\,e^{-|y|/\ell_2}\,
K_0\!\left(\sqrt{\frac{r}{D}}\sqrt{x_1^2+(x_2-y)^2}\right).
\end{aligned}
\end{equation}
In the following sections, we show that these one- and two-point marginals can always be obtained for arbitrary \(m\) and \(N\), while the full stationary state takes a particularly simple form only in the cases $m=1$ and $m=N-1$, in addition to the known case $m=N$ \cite{Biroli2023}. 
Here, with the specific choice $N=3$ and $m=2$, the JPDF in Fourier space can be obtained by inserting \eqref{eq:N3_P1tilde} into \eqref{eq:N3_statstate_kspace}:
\begin{equation}\label{eq:Ptilde_factorized_APP}
\tilde P_{m=2}(k_1,k_2,k_3)
= \frac{1}{3}\,\frac{r}{r+D(k_1^2+k_2^2+k_3^2)}
\sum_{j=1}^3
\frac{1}{1+\ell_2^2 k_j^2}\;,
\qquad
\ell_2=\sqrt{\frac{3D}{2r}}.
\end{equation}
We now notice that
\begin{equation}\label{eq:N3_Psimultreset}
\mathcal{F}^{-1}\!\left[\frac{1}{1+\ell^2\sum_{i=1}^3 k_i^2}\right]
=\frac{1}{\ell^3}\,\frac{1}{(2\pi)^{3/2}}
\left(\frac{|\mathbf{x}|}{\ell}\right)^{-\nu}
K_{\nu}\!\left(\frac{|\mathbf{x}|}{\ell}\right)=P_{m=3}(\mathbf{x}),
\quad
\ell=\sqrt{\frac{D}{r}},\quad
\nu=\frac{1}{2},
\end{equation}
where \(\mathcal{F}^{-1}\) denotes the inverse FT, $|\mathbf{x}|= \sqrt{x_1^2+x_2^2+x_3^2}$ and \(K_{\nu}(z)\) is the modified Bessel function of the second kind of order \(\nu\).
This expression is precisely the stationary state $P_{m=3}(\mathbf{x})$ of a system of \(N=3\) particles that are all simultaneously reset (\(m=3\)) \cite{Biroli2023}.
We can now use the identity
\begin{equation}
    K_{1/2}(z)=\sqrt{\frac{\pi}{2z}}\, e^{-z}
\end{equation}
to rewrite Eq.~\eqref{eq:N3_Psimultreset} as
\begin{equation}
    P_{m=3}(\mathbf{x})
    =\frac{1}{\ell^3}\,\frac{1}{4\pi}
    \left(\frac{|\mathbf{x}|}{\ell}\right)^{-1}
    e^{-|\mathbf{x}|/\ell}.
\end{equation}
In addition, the sum in ~\eqref{eq:Ptilde_factorized_APP} contains terms that are exactly the one-point marginal in Fourier space $\tilde{P}_{m=2}^1(k_j)$, and therefore their inverse FT gives \eqref{eq:N3_P1tilde}.
Using these observations and the convolution property of the FT, the JPDF can be written as
\begin{equation}\label{eq:SSfinal_3particles}
\begin{aligned}
P_{m=2}(x_1,x_2,x_3)
&= \frac{1}{3}
\int_{-\infty}^{\infty}\dd y\,P_{m=2}^1(x_1-y)\,
P_{m=3}\left(y,x_2,x_3\right)\\
&\quad +\frac{1}{3}
\int_{-\infty}^{\infty}\dd y\,P_{m=2}^1(x_2-y)\,
P_{m=3}\left(x_1,y,x_3\right)\\
&\quad +\frac{1}{3}
\int_{-\infty}^{\infty}\dd y\,P_{m=2}^1(x_3-y)\,
P_{m=3}\left(x_1,x_2,y\right).
\end{aligned}
\end{equation}
This shows that, for \(N=3\) and \(m=2\), the stationary JPDF can be written as a sum of convolutions between the one-point marginal \(P^{1}_{m=2}\) and the fully synchronous JPDF \(P_{m=3}\). Physically, the system behaves as if the dynamics were fully synchronous, except that at each resetting event one particle is effectively replaced by an independent particle drawn from its stationary one-body distribution. All three possible choices of the pair of particles to be reset occur with equal probability, resulting in the prefactor \(1/3\). This construction naturally generalizes to arbitrary \(N\) in the case \(m = N-1\), as expressed in Eq.~\eqref{eq:PNmeno1_SS} below.

\section{Fokker--Planck equation for arbitrary $m$ and $N$}
\label{sec:FP_derivation}

In this section, we show that the structure uncovered in the illustrative case \(N=3\) and \(m=2\) naturally extends to arbitrary values of \(m\) and \(N\).
We now consider the JPDF $P_m(\mathbf{x},t)$, where $\mathbf{x}=\{x_1,...,x_N\}$ are the positions of the \(N\) particles.
Each particle performs an independent Brownian motion with diffusion constant \(D\), and, at Poissonian times with rate \(r\), a subset of exactly \(m\) particles is reset to the origin.
We define the $n$-point marginals of the JPDF in direct and Fourier space, respectively as
\begin{equation}\label{eq:marginal_DEF}
\begin{aligned}
    P_m^{n}(x_1,\ldots,x_n,t)
    &=\!\int\!\!\int_{-\infty}^{\infty}\!\!dx_{n+1}...dx_N\,P_m(\mathbf{x},t),
    \\[6pt]
    \tilde P_m^{n}(k_1,\ldots,k_n,t)
    &=\!\int\!\!\int_{-\infty}^{\infty}\!\!dx_1... dx_n\,e^{i(k_1x_1+\cdots+k_nx_n)}\,P_m^{n}(x_1,\ldots,x_n,t).
\end{aligned}
\end{equation}
The time evolution of \(P_m(\mathbf{x},t)\) is governed by the Fokker--Planck equation
\begin{equation}\label{eq:FPeq_xspace_APP}
\partial_t P_m(\mathbf{x},t)
= D \sum_{i=1}^{N} \partial_{x_i}^2 P_m(\mathbf{x},t)
- r\, P_m(\mathbf{x},t)
+ \frac{r}{\binom{N}{m}}
\sum_{\substack{S_m}}\,
\prod_{i \in S_m}\! \delta(x_i)\,\,
P_m^{N-m}(\mathbf{x}_{S_m^{\mathrm{c}}},t)
\end{equation}
where
\begin{equation}\label{eq:marginal_Nminm}
    P_m^{N-m}(\mathbf{x}_{S_m^{\mathrm{c}}},t)=\int \!\!\int d \mathbf{x}_{S_m}  \,
    P_m(\mathbf{x},t)\,
\end{equation}
is the $(N-m)$-point marginal, as defined in \eqref{eq:marginal_DEF}. The notation $\int \!\!\int d \mathbf{x}_{S_m}$ indicates that we are integrating over all the components $x_i$ such that $i\in S_m$, where $S_m$ is a subset $S_m\subseteq\{ 1,...,N\}$ such that $|S_m|=m$. It follows that $P_m^{N-m}$ only depends on the components $x_i$ of $\mathbf{x}$ such that $i\notin S_m$, or, equivalently, such that $i\in S_m^c$ where $S_m^c$ is the set $S_m^c=\{1, \dots, N\} \setminus S_m$.We compactly denote the remaining components as $\mathbf{x}_{S_m^{\mathrm{c}}}\in \mathbb{R}^{N-m}$.
To be clearer, we consider again the case $N=3$ and $m=2$. This means $P_m^{N-m}\to P_2^1$ and $S_m\to S_2$. In this case, there are three possibilities for the set $S_2$. For example we may have $S_m=\{1,2\}$, which implies $S_m^c=\{3\}$. Then \eqref{eq:marginal_Nminm} becomes
\begin{equation}
    P_2^{1}(\mathbf{x}_{S_2^{\mathrm{c}}}=x_3,t)=\int \!\!\int d \mathbf{x}_{S_2}  \,
    P_2(\mathbf{x},t)=\int \!\!\int d x_1 dx_2  \,
    P_2(\mathbf{x},t)
\end{equation}
With the choices $S_2=\{2,3\}$ and $S_2=\{1,3\}$ we obtain instead, respectively, $P_2^{1}(\mathbf{x}_{S_2^{\mathrm{c}}}=x_1,t)$ and $P_2^{1}(\mathbf{x}_{S_2^{\mathrm{c}}}=x_2,t)$. 
The sum $\sum_{\substack{S_m}}$ runs over all the possible subsets $S_m$.
Since all particles play the same role in the dynamics, the three functions $P_2^{1}(x_1,t)$, $P_2^{1}(x_2,t)$ and  $P_2^{1}(x_3,t)$ all have the same functional form.

We now analyze the terms appearing in the right-hand side of \eqref{eq:FPeq_xspace_APP}.
The first term represents standard diffusion with diffusion constant $D$.
The second term accounts for the loss of probability at position $\mathbf{x}$ due to the resetting events occurring at rate $r$. Whenever a resetting happens, some of the particles currently at $\mathbf{x}$ are reset to the origin, thus reducing the probability at that point.
The last term captures the effect of batch resetting, and it generalizes the terms in the square bracket of \eqref{eq:FP_x_N3_m2}.
Each time a reset occurs, a random subset $S_m\subseteq\{ 1,...,N\}$ made of $|S_m|=m$ particles is selected from the $N$ particles. This selection is uniform, meaning that any subset of size $m$ is equally likely to be chosen. The contribution to the probability density from resetting these $m$ particles is given by $\prod_{i \in S_m} \delta(x_i)$, where the Dirac delta $\delta(x_i)$ forces the positions of the selected particles to reset to the origin.
The remaining $N-m$ particles are unaffected by the resetting event and continue to evolve according to their diffusion dynamics. Therefore, we must multiply by the marginal distribution of the positions of these $N-m$ particles, namely $P_m^{N-m}(\mathbf{x}_{S_m^{\mathrm{c}}},t)$. 
The factor $1/\binom{N}{m}$ is the probability of selecting a particular subset, since any subset $S$ of size $m$ is equally probable. We then sum over all possible subsets $S_m$ to account for all possible choices of particles to reset.
Finally, we multiply the probability of a resetting event occurring per unit time, which is simply the rate $r$.

It is now convenient to take a FT of \eqref{eq:FPeq_xspace_APP}, which leads to
\begin{equation}\label{eq:FP_inkspace_APP}
\partial_t \tilde{P}_m(\mathbf{k},t)
= -\big( D k^2 + r \big)\, \tilde{P}_m(\mathbf{k},t)
+ \frac{r}{\binom{N}{m}}
\sum_{\substack{S_m}}
\tilde{P}_m^{N-m}\!\big(\mathbf{k}_{S_m^{\mathrm{c}}}, t\big) \,,
\end{equation}
where we have defined 
\begin{equation}\label{eq:definiton_FT_SM}
    \tilde{P}_m(\mathbf{k},t)=\!\int\!\! \int d \mathbf{x} \, e^{i \mathbf{k}\cdot\mathbf{x}} \, P_m(\mathbf{x},t).
\end{equation}
and $k^2 = |\mathbf{k}|^2 = k_1^2 + \cdots + k_N^2$.
The function  
\(\tilde{P}_m^{N-m}\!\big(\mathbf{k}_{S_m^{\mathrm{c}}}, t\big)\)  
denotes the \((N-m)\)-dimensional FT of the marginal distribution  
\(P_m^{N-m}(\mathbf{x}_{S_m^{\mathrm{c}}},t)\).
For instance, when \(N=3\) and \(m=2\), one possible choice is \(S_2=\{1,2\}\), which implies
\(S_2^{\mathrm{c}}=\{3\}\). In this case, \(\mathbf{k}_{S_2^{\mathrm{c}}}=(k_3)\) and
\begin{equation}
\tilde{P}_m^{N-m}\!\big(\mathbf{k}_{S_m^{\mathrm{c}}},t\big)
=\tilde{P}_2^{1}(k_3,t)
=\int_{-\infty}^{\infty} dx_3\, e^{i k_3 x_3}\, P_2^{1}(x_3,t).
\end{equation}
It is important to note that the Fourier-space marginal \(\tilde{P}_2^{1}(k_3,t)\) can
equivalently be obtained from the full Fourier-transformed JPDF
\(\tilde{P}_2(\mathbf{k},t)\) by evaluating it at \(\mathbf{k}=(0,0,k_3)\), or, by symmetry,
at \((k_3,0,0)\) or \((0,k_3,0)\).
This follows directly from the definition~\eqref{eq:definiton_FT_SM}: setting a component
\(k_i=0\) corresponds to marginalizing the direct space JPDF over the associated spatial coordinate \(x_i\).
The same argument extends to higher-order marginals: by setting \(n\) components of
\(\mathbf{k}\) to zero in the full Fourier-transformed JPDF, one directly obtains the FT of the corresponding \((N-n)\)-point marginal.
This observation will prove useful in the following sections.
Finally, one can readily check that setting \(N=3\) and \(m=2\) in \eqref{eq:FP_inkspace_APP} reproduces \eqref{eq:FP_k_N3_m2_BIS}.

\section{One- and two-point marginals}
\label{sec:Oneandtwopointmarginals}

From \eqref{eq:FP_inkspace_APP} we can derive an expression for the stationary state of the system. By setting the temporal derivative to zero, we obtain:
\begin{equation}\label{eq:SSP(k)_APP}
\tilde{P}_m(\mathbf{k}) = \frac{ \displaystyle \frac{r}{\binom{N}{m}}
        \sum_{\substack{S_m}}
        \tilde{P}_m^{N-m}\!\big(\mathbf{k}_{S_m^{\mathrm{c}}}\big)}
     { \displaystyle r + D \sum_{i=1}^{N} k_i^{2} } \, .
\end{equation}
This equation can, in principle, be used to determine the full JPDF of the system and its marginals for all values of \( m \).
Specifically, since \eqref{eq:SSP(k)_APP} holds for any vector $\mathbf{k}$, we can evaluate it at any convenient value. 
Using the observation made above that setting components of $\mathbf{k}$ to zero in $\tilde{P}_m(\mathbf{k})$ yields Fourier-space marginals, we may choose $\mathbf{k}=(k,0,\dots,0)$.
Doing this, the left-hand side becomes just the FT \( \tilde{P}_m^1(k) \) of the one-point marginal defined in \eqref{eq:marginal_DEF}. In this case, the right-hand side will depend only on \( \tilde{P}_m^1(k) \), allowing us to derive the following expression:
\begin{equation}\label{eq:P1tilde_APP}
\tilde{P}_m^1(k) = \frac{1}{1+\ell_m^2 k^2}, \qquad \tilde{r} = \frac{m}{N}r, \qquad  \ell_m^2 = \frac{D}{\tilde{r}}.
\end{equation}
Next, we apply the same reasoning to derive the two-point marginal. We evaluate \eqref{eq:SSP(k)_APP} at $\mathbf{k} = (k_1, k_2, 0, \ldots, 0)$.
With this choice, the left-hand side of \eqref{eq:SSP(k)_APP} reduces to the
FT of the two-point marginal, $\tilde{P}_m^{2}(k_1,k_2)$.
On the right-hand side, the sum over subsets $S_m$ produces contributions that
depend on how the indices $1$ and $2$ are distributed with respect to $S_m$.
Specifically, depending on whether neither of the indices $1,2$ belongs to $S_m$,
exactly one of them belongs to $S_m$, or both belong to $S_m$, the corresponding
terms are respectively proportional to $\tilde{P}_m^{2}(k_1,k_2)$, to
$\tilde{P}_m^{1}(k_1)$ or $\tilde{P}_m^{1}(k_2)$, or to a constant.
This classification allows us to express \eqref{eq:SSP(k)_APP} as a closed
equation involving only the known one-point marginal and the unknown two-point
marginal, from which $\tilde{P}_m^{2}(k_1,k_2)$ can be obtained explicitly as
\begin{equation}\label{eq:P2tilde_APP}
\tilde{P}_m^2(k_1,k_2) = \frac{\,(2-\beta) + (\beta-1) \left[\tilde{P}_m^1(k_1) + \tilde{P}_m^1(k_2)\right]}{ \,  \, \beta + \ell_m^2(k_1^2 + k_2^2)}, \qquad \beta=\frac{2N-m-1}{N-1}.
\end{equation}
In principle, repeating this argument allows one to obtain all the marginals up to the full JPDF in Fourier space. However, their expressions quickly become complex for marginals involving more than two particles.
The FT in Equations \eqref{eq:P1tilde_APP} and \eqref{eq:P2tilde_APP} can be inverted to obtain:
\begin{eqnarray}\label{eq:1-point_marg_APP}
    P_m^1(x) = \frac{1}{\ell_m} \mathcal{R}\left( \frac{x}{\ell_m} \right), \quad \mathcal{R}(u) = \frac{1}{2}e^{-|u|},
\end{eqnarray}
and 
\begin{equation}\label{eq:2-point_marg_APP}
P_m^2(x_1, x_2)
= (2-\beta) \, h(x_1, x_2)
+ (\beta-1) \left[
    I(x_1, x_2)
  + I(x_2, x_1)
  \right],
\end{equation}
where
\begin{align}
h(x_1, x_2) &= \frac{1}{2\pi \ell_m^2} \, K_0\!\left( \sqrt{\beta}\, \sqrt{\left(\frac{x_1}{\ell_m}\right)^2 + \left(\frac{x_2}{\ell_m}\right)^2} \right), \qquad \ell_m=\sqrt{\frac{D}{\tilde{r}}},\\[6pt]
I(x_1, x_2) &= (P_m^1 *_1 h)(x_1, x_2) = \int_{-\infty}^{\infty} \frac{dy}{2\ell_m} \, e^{-|y|/\ell_m} \,
    h(x_1 - y, x_2),
\end{align}
and \( K_0(z) \) is the modified Bessel function of the second kind of order zero. The symbol $*_1$ indicates a convolution over the first coordinate.
It is useful to observe that the parameters $\beta-1$ and $2-\beta$ admit a simple probabilistic interpretation.
Fix two particles $i$ and $j$ and condition on particle $i$ being reset during a given event.
Since the remaining $m-1$ particles are chosen uniformly among the other $N-1$, the probability that particle $j$ is also reset is
\[
2-\beta=\frac{m-1}{N-1},
\]
while the complementary probability that particle $j$ is not reset is
\[
\beta-1=\frac{N-m}{N-1}.
\]
These two probabilities sum to unity and directly determine the relative weights of the different contributions in \eqref{eq:P2tilde_APP}.
In particular, the coefficient $2-\beta$ multiplies the contribution associated with the simultaneous resetting of both particles; accordingly, the function $h(x_1,x_2)$ has the same functional form as the stationary joint distribution in the fully synchronous case~\cite{Biroli2023}.
Conversely, the coefficient $\beta-1$ weighs the contributions arising from events in which only one of the two particles is reset, which can be interpreted similarly to \eqref{eq:SSfinal_3particles}.

We can rewrite the function \( P_m^2(x_1, x_2) \) in the scaling form
\begin{equation}\label{eq:Pstrana_APP}
    P_m^2(x_1, x_2) = \frac{1}{\ell_m^2} \mathcal{P}_{\beta}\left(\frac{x_1}{\ell_m}, \frac{x_2}{\ell_m}\right),
\end{equation}
where the scaling function is given by
\begin{equation}\label{eq:P(u,v)_APP}
\begin{split}
\mathcal{P}_{\beta}(u,v) &= \frac{2-\beta}{2\pi} K_0\!\left( \sqrt{\beta} \sqrt{u^2 + v^2} \right)  \\
&\quad + \frac{\beta-1}{4\pi} \Bigg[
    \int_{-\infty}^{\infty} \! dz\, e^{-|z|} K_0\!\left( \sqrt{\beta} \sqrt{(u-z)^2 + v^2} \right) \\
&\qquad\qquad
   + \int_{-\infty}^{\infty} \! dz\, e^{-|z|} K_0\!\left( \sqrt{\beta} \sqrt{u^2 + (v-z)^2} \right)
\Bigg].
\end{split}
\end{equation}
The function $\mathcal{P}_{\beta}(0,v)$ behaves as
\begin{equation}
    \mathcal{P}_{\beta}(0,v)\simeq
    \begin{cases}
        -\frac{2-\beta}{2\pi} \, \ln \left(\frac{\sqrt{\beta}}{2}|v|\right), & v\to 0,\\
        \frac{\sqrt{\beta-1}}{4} \,e^{-v}, & v\to \infty.
    \end{cases}
\end{equation}
Integrating over one of the two variables, we recover the one-point marginal
\begin{equation}
    \int dv \, \mathcal{P}_{\beta}(u,v) = \mathcal{R}(u)=\frac{1}{2}e^{-|u|}.
\end{equation}
Setting $m=N$, which means $\beta=1$, we recover the known result for $N$ particles that are all reset simultaneously \cite{Biroli2023}:
\begin{equation}
    \mathcal{P}_{\beta}(u,v)=\frac{1}{2\pi}K_0\left( \sqrt{u^2+v^2} \right).
\end{equation}
We observe that the scaling function $\mathcal{R}(u)$ does not depend on $m$ for any value of $N$. 
However, the situation changes for the joint function $\mathcal{P}_{\beta}(u,v)$: as soon as $m \neq N$ ($\beta \neq 1$), additional terms appear in \eqref{eq:P(u,v)_APP}. 
This implies that one-particle observables, such as the density profile, retain the same qualitative form as in the fully synchronous case, whereas any observable involving two particles is qualitatively modified when $m \neq N$. This is precisely what happens for the two-point correlators studied below and in the Letter.

\section{JPDF for $m=N-1$}\label{sec:JPDF_m=Nminus1_APP}
For the case $m=N-1$, it is possible to obtain an explicit expression of the full stationary state $P_{N-1}(\mathbf{x})$, and not only for its first two marginals. Starting from \eqref{eq:SSP(k)_APP}, one can directly set $m=N-1$, which leads to an expression involving only the one-point marginal. Since we have already found the one-point marginal in general for any $m$ (see \eqref{eq:1-point_marg_APP}) we get 
\begin{equation}\label{eq:tildeP_nmeno1_APP}
    \tilde{P}_{N-1}(\mathbf{k})
= \frac{r}{D \sum_{i=1}^{N} k_i^{2} + r}\;
  \frac{1}{N}\sum_{j=1}^{N}
  \frac{\tilde{r}}{D\,k_j^{2} + \tilde{r}}, \qquad m=N-1,
\end{equation}
where $\tilde{r}=\frac{N-1}{N}r$.
To invert the FT in \eqref{eq:tildeP_nmeno1_APP}, we notice that

\begin{equation}\label{eq:PmequalN}
\mathcal{F}^{-1}\!\Big[\frac{r}{D\sum_i k_i^2+r}\Big]
=\frac{1}{\ell^N}\, \frac{1}{(2\pi)^{N/2}} \left( \frac{|\mathbf{x}|}{\ell} \right)^{-\nu} \, K_{\nu}\left( \frac{|\mathbf{x}|}{\ell} \right)=P_{m=N}(\mathbf{x}),
\end{equation}
where $\mathcal{F}^{-1}$ indicates the inverse FT, $K_{\nu}(z)$ is the modified Bessel function of the second kind of order $\nu$, \(|\mathbf{x}|=\sqrt{x_1^2+...+x_N^2}\), \(\ell=\sqrt{D/r}\) and \(\nu=\tfrac{N}{2}-1\). 
The function $P_{m=N}(\mathbf{x})$ is exactly the JPDF of a batch resetting system of $N$ particles which are all simultaneously reset (i.e. with $m=N$)\cite{Biroli2023}.
We also define
\begin{equation}\label{eq:F2x}
    F_j(\mathbf{x})=\mathcal{F}^{-1}\!\Big[\frac{\tilde r}{D k_j^2+\tilde r}\Big]
=\frac{1}{2\ell_{N-1}}\,e^{-|x_j|/\ell_{N-1}} \prod_{i\neq j}\delta(x_i)=\frac{1}{\ell_{N-1}}\mathcal{R}\left( \frac{|x_j|}{\ell_{N-1}} \right)\prod_{i\neq j}\delta(x_i),
\qquad
\ell_{N-1}=\sqrt{\frac{ND}{(N-1)\, r}}.
\end{equation}
Interestingly, in the above expression appears the one-point marginal (i.e. the average density of particles) $P_{m=N-1}^1(x_j)=\frac{1}{\ell_{N-1}}\mathcal{R}\left( \frac{|x_j|}{\ell_{N-1}}\right)$ of our system with $m=N-1$ (see \eqref{eq:1-point_marg_APP}).
Equations \eqref{eq:PmequalN} and \eqref{eq:F2x}, together with the convolution property of the FT, allow us to express the JPDF $P_{N-1}(\mathbf{x})$ as
\begin{equation}\label{eq:PNmeno1_SS}
    P_{N-1}(\mathbf{x})=\frac{1}{N}\sum_{j=1}^N\big(P_{m=N} * F_j\big)(\mathbf{x})
=\frac{1}{N}\sum_{j=1}^N
\int_{-\infty}^{\infty}\!dy\;
P_{m=N}\!\big(x_1,\ldots,x_{j-1},y,x_{j+1},\ldots,x_N\big)\;
\frac{1}{\ell_{N-1}}\,\mathcal{R}\left( \frac{|x_j-y|}{\ell_{N-1}}\right),
\end{equation}
where the symbol $*$ indicates a convolution.
The interpretation of this expression is straightforward. At every reset event, all particles except one are brought back to the origin. As a result, the stationary state can be constructed by taking the stationary JPDF $P_{m=N}(\mathbf{x})$, corresponding to the fully simultaneous resetting, and replacing the coordinate 
of the particle that is not reset with a position drawn from its one-body stationary distribution $\frac{1}{\ell_{N-1}}\,\mathcal{R}\left( \frac{|x_j-y|}{\ell_{N-1}}\right)$.
Since any of the $N$ particles can play this role with equal probability, the final stationary distribution is obtained by averaging this construction over all possible choices of that particle.

\section{JPDF for $m=1$}\label{sec:JPDF_m=1_APP}

In the case $m=1$, each time a reset event occurs only a single particle is reset, while the remaining $N-1$ continue their diffusive motion unperturbed. As no two particles are ever reset simultaneously, the sole mechanism capable of generating correlations is absent, since there are no direct interactions between the particles. Consequently, the JPDF must factorize into a product of single-particle distributions. Because all particles play the same role in the system, each one-point marginal must coincide with the average particle density. The JPDF therefore takes the form
\begin{equation}\label{eq:JPDF_m=1_APP}
    P_{m=1}(\mathbf{x})=\prod_{i=1}^N P_{m=1}^1(x_i),
\end{equation}
where $P_{m=1}^1(x)$ is the one-point marginal given in \eqref{eq:1-point_marg_APP}. Explicitly,
\begin{equation}\label{eq:JPDF_m=1explicit_APP}
    P_{m=1}(\mathbf{x})
    =\prod_{i=1}^N \frac{1}{2\,\ell_1} e^{-|x_i|/\ell_1},
    \qquad \ell_1=\sqrt{\frac{ND}{r}}.
\end{equation}
One may also take the FT of \eqref{eq:JPDF_m=1explicit_APP} and check that it is solution of \eqref{eq:SSP(k)_APP}.

\section{Correlators: $\mathcal{C}_2(t)$ and $a(t;m,N)$}
\label{sec:correlators}

As discussed in the main text, the flip symmetry \(x_i \to -x_i\) implies that the first non-trivial correlations of a pair of particles in the system are captured by the \(2\times 2\) correlation matrix \({\mathbf{C}}\), defined by  
\begin{equation}\label{eq:defdiCij}
    C_{ij}(t)= \langle x_i^2(t) x_j^2(t) \rangle - \langle x_i^2(t) \rangle\,\langle x_j^2(t) \rangle, 
    \qquad i,j\in\{1,2\}.
\end{equation}
Permutation symmetry among the particles immediately yields  
\begin{align}\label{eq:quantities}
    C_{ii}(t) &= \mathcal{C}_1(t)
    = \langle x_i^4(t) \rangle - \langle x_i^2(t) \rangle^2,\\[2mm]
    C_{ij}(t) &= \mathcal{C}_2(t)
    = \langle x_i^2(t) x_j^2(t) \rangle - \langle x_i^2(t) \rangle\,\langle x_j^2(t) \rangle,
    \qquad i\neq j.
\end{align}
We then introduce the normalized correlation coefficient  
\begin{equation}\label{eq:a_C2/C1_rewritten}
    a(t;m,N) = \frac{\mathcal{C}_2(t)}{\mathcal{C}_1(t)}.
\end{equation}
Equivalently, since $\mathbf{C}$ is symmetric and therefore admits two real eigenvalues (which we denote by $\lambda_+(t) > \lambda_-(t)$), one finds
\begin{equation}\label{eq:a_eigvals_rewritten}
    a(t;m,N) = \frac{\lambda_+(t)-\lambda_-(t)}{\lambda_+(t)+\lambda_-(t)}.
\end{equation}
The coefficient \(a(t;m,N)\) lies in the interval \([0,1]\): it vanishes when \(x_i^2\) and \(x_j^2\) are uncorrelated, and reaches unity when they are perfectly correlated.

The quantities in \eqref{eq:quantities} and \eqref{eq:a_C2/C1_rewritten} can be computed in two equivalent ways. A first approach consists of multiplying eq.~\eqref{eq:FPeq_xspace_APP} by an arbitrary observable \( f(\mathbf{x}) \) and integrating over all configurations \(\mathbf{x}\). This yields a differential equation for the expectation value \(\langle f(\mathbf{x}) \rangle\), which can then be explicitly solved for observables such as \(f(\mathbf{x}) = x_i^2 x_j^2\), \(x_i^2\), or \(x_i^4\).  
Alternatively, one can work directly in Fourier space. From \eqref{eq:FP_inkspace_APP}, with the same reasoning used to derive \eqref{eq:P1tilde_APP} and \eqref{eq:P2tilde_APP}, one can derive the expressions for the full time-dependent solutions of the marginals $\tilde{P}_1(k,t)$ and $\tilde{P}_2(k_1,k_2,t)$. The results are:
\begin{equation}\label{eq:tildeP1time_APP}
\tilde{P}_m^1(k,t) =\frac{\tilde{r}}{\tilde{r}+Dk^2} + \frac{Dk^2}{Dk^2+\tilde{r}}\, e^{-(Dk^2+\tilde{r})t}
\end{equation}
and 
\begin{equation}\label{eq:tildeP2(k_1,k_2,t)_APP}
    \tilde{P}_m^2(k_1,k_2,t)=e^{-\lambda_{2} t}
+ \frac{\tilde{r}\,(2-\beta)}{\lambda_2}\bigl(1 - e^{-\lambda_{2} t}\bigr)
+ \tilde{r}\, (\beta-1)\,\sum_{j=1}^2\!\left\{
\frac{\tilde{r}}{\gamma(k_{j})}\,\frac{1 - e^{-\lambda_{2} t}}{\lambda_{2}}
+ \left(1 - \frac{\tilde{r}}{\gamma(k_{j})}\right)\!
\frac{e^{-\gamma(k_{j}) t} - e^{-\lambda_{2} t}}{\,\lambda_{2} - \gamma(k_{j})\,}
\right\}
\end{equation}
with 
\[
\lambda_{2} = D(k_{1}^{2} + k_{2}^{2}) + \tilde{r}\beta, 
\qquad
\gamma(k)=Dk^2+\tilde{r}.
\]
By differentiating Eqs.~\eqref{eq:tildeP1time_APP} and \eqref{eq:tildeP2(k_1,k_2,t)_APP} the desired moments can be obtained through the identities:
\begin{equation}
\begin{aligned}
\langle x_i^2(t) \rangle = -\left.\frac{\partial^2 \tilde{P}_m^1(k,t)}{\partial k^2}\right|_{k=0}, \quad
\langle x_i^4(t) \rangle  = \left.\frac{\partial^4 \tilde{P}_m^1(k,t)}{\partial k^4}\right|_{k=0}, \quad
\langle x_i^2 x_j^2 (t)\rangle = \left.\frac{\partial^2}{\partial k_1^2}\frac{\partial^2}{\partial k_2^2}\tilde{P}_m^2(k_1,k_2,t)\right|_{k_1,k_2=0}.
\end{aligned}
\end{equation}
It is then straightforward to get
\begin{align}
\mathcal{C}_1(t) &= \frac{4D^2}{\tilde{r}^2}\, f_1(\tilde{r}t),
&& f_1(z)= 5 - 4e^{-z} - 6z e^{-z} - e^{-2z},
\label{eq:C1_APP} \\[4pt]
\mathcal{C}_2(t) &= \frac{4D^2}{\tilde{r}^2}\, f_2(\tilde{r}t;\beta),
&& f_2(z;\beta)=2\!\left(\frac{1-e^{-\beta z}}{\beta}
      -\frac{e^{-z}-e^{-\beta z}}{\beta-1}\right)
      -1 - e^{-2z} + 2e^{-z},
\label{eq:C2_APP} \\[4pt]
a(t;m,N) &= A(\tilde{r}t;\beta),
&& A(z;\beta)=\frac{f_2(z;\beta)}{f_1(z)}.
\label{eq:a_APP}
\end{align}
where $\tilde{r}=\frac{m}{N}r$ and $\beta=\frac{2N-m-1}{N-1}$. We recall here that $\beta \in [1,2]$, where $\beta=1$ corresponds to the classical case with $m=N$, while $\beta=2$ corresponds to $m=1$. In the fully synchronous limit $\beta\to 1$ ($m\to N$), the function $f_2(z;\beta)$ converges to the known result
$f_2(z;1)=1-2ze^{-z}-e^{-2z}$~\cite{DeMauro2025}.
The asymptotics of these three functions, when $\beta>1$, are 
\begin{equation}
    f_1(z)\approx
    \begin{cases}
        2z^2, & z\to0,\\
        5-6ze^{-z}, & z\to \infty.
    \end{cases}
\end{equation}

\begin{eqnarray}
    f_2(z;\beta)\approx 
    \begin{cases}
        \dfrac{2-\beta}{3}\, z^3, & z\to0,\\[6pt]
        \left( \dfrac{2}{\beta}-1 \right)-2\dfrac{2-\beta}{\beta-1}\, e^{-z}, & z\to \infty.
    \end{cases}
\end{eqnarray}

\begin{equation}\label{eq:asymptotics_A(z;beta)}
A(z;\beta)\approx
\begin{cases}
\displaystyle 
\dfrac{2-\beta}{6}z, & z\to 0^+,\\[8pt]
\displaystyle \dfrac{1}{5} \left( \dfrac{2}{\beta}-1 \right)
+\Bigg[\dfrac{6(2-\beta)}{25\,\beta}\,z
+\dfrac{2(3\beta^{2}-4\beta-4)}{25\,\beta(\beta-1)}\Bigg]e^{-z}, & z\to\infty,\, \beta>1.
\end{cases}
\end{equation}

\subsection{Analysis of the maximum $z^*(\beta)$ of $A(z;\beta)$}\label{sec:analysismaxd=1}

From Fig. 3 of the main Letter and Eq. \eqref{eq:asymptotics_A(z;beta)}, one sees that \(A(z;\beta)\) must have a maximum at some finite value $z^*(\beta)$ for any $1<\beta<2$. However, a closed analytical expression for \(z^*(\beta)\) is difficult to obtain. Nevertheless, we can make progress by considering the regime \(\beta=1+\epsilon\) with $\epsilon = \frac{N-m}{\,N-1\,}$
and by searching for an expansion of \(z^*(\epsilon)\) for \(\epsilon\to0\) (i.e.\ \(m\to N,\) with $N$ large).
A numerical analysis shows (see Fig. 4 in the main Letter) that \(z^*(\epsilon)\to\infty\) as \(\epsilon\to 0\). Therefore, when analyzing the extremum condition \(\partial_z A(z;\beta) = F(z,\beta)=0\), we must take the joint limit \(z\to\infty\), \(\epsilon\to0\) while keeping the product \(z\epsilon\) finite. Expanding \(F(z,1+\epsilon)\) in this double scaling limit yields, at leading order,
\begin{eqnarray}\label{eq:F=0_APP}
    F(\beta=1+\epsilon,z)\approx\frac{2}{25} e^{-z\epsilon}e^{-z}\,\frac{
e^{z\epsilon}(\epsilon-1)\big[-6(1+\epsilon)+3\epsilon z+1\big]-5(1+\epsilon)}{\epsilon(1+\epsilon)}.
\end{eqnarray}
We now introduce the asymptotic ansatz
\[
z^*(\epsilon)=\frac{a_0}{\epsilon}+a_1,
\]
substitute it into \eqref{eq:F=0_APP}, expand for small $\epsilon$, and set the coefficients of the various orders in $\epsilon$ to zero. This yields the coupled equations
\[
e^{a_0}(5 - 3a_0)=5,
\qquad
e^{a_0}\big[(5 - 3a_0)(a_1 - 1) + 6 - 3a_1\big]=5.
\]
The first equation also admits the trivial solution $a_0=0$. Discarding it, the only nontrivial solution is $a_0=1.1262\ldots$, and the second equation then gives $a_1=2$.
Thus, for \(\epsilon\to0\) the position of the maximum diverges as  
\begin{equation}\label{eq:zstar(eps)_APP}
    z^*(\epsilon)\simeq \frac{1.1262}{\epsilon}+2.
\end{equation}
This expansion is plotted in Fig. 4 in the main Letter together with the results of numerical locations of the maximum. For $\epsilon=0$ exactly, the function $A(z;\beta=1)$ is instead strictly monotonic in $z$. 

We can also analyze the behavior of \( z^{*}(\beta) \) in the limit \( \beta \to 2^{-} \) (i.e. $N\to \infty$ and $m>1$). In this regime, the amplitude of the scaling function \(A(z;\beta)\) vanishes, as correlations disappear when particles are never reset together. Nevertheless, the position of the maximum remains finite:
\begin{equation}\label{eq:zbar}
    z^{*}(\beta\to 2^-) \longrightarrow \bar z = 2.6495\dots .
\end{equation}
Thus, throughout the whole interval \(1 < \beta < 2\), the decorrelation mechanism generated by batch resetting is characterized by a timescale which is always greater than $\bar z$. The function \(z^{*}(\beta)\) interpolates between the finite limit \(\bar z\) at \(\beta\to 2^{-}\) and the divergent behavior obtained as \(\beta\to 1^+\).
The result in \eqref{eq:zbar} can be derived by setting \(\beta = 2 - \delta\) with $\delta=\frac{m-1}{N-1}$ in \eqref{eq:a_APP}, expanding to first order in \(\delta\), differentiating with respect to \(z\), and imposing the extremum condition. This leads to the equation
\begin{equation}
6 e^z z^2 + 13 e^z z - 10 e^{2 z} z - 3 e^{3 z} z + 13 e^z - 23 e^{2 z} + 11 e^{3 z} - 1 = 0.
\end{equation}
This equation admits a unique positive solution, $\bar z = 2.6495\ldots$.

\subsection{Stationary correlators}

We have already obtained the full time dependence of the second-order correlations between a pair of particles, namely 
\begin{equation}\label{eq:C2definition}
    \mathcal{C}_2(t)
    = \langle x_i^2(t) x_j^2(t) \rangle - \langle x_i^2(t) \rangle\,\langle x_j^2(t) \rangle,
    \qquad i\neq j.
\end{equation}
This function depends on the number of particles $N$ and the number of reset particles $m$; its explicit expression is given in \eqref{eq:C2_APP}. 
For notational simplicity, we omit the explicit dependence on $m$ and $N$ in \eqref{eq:C2definition}.
Taking the limit $t\to \infty$ in \eqref{eq:C2_APP} and using $\beta=\frac{2N-m-1}{N-1}$ we get the stationary correlator
\begin{equation}\label{eq:V(m,N)_d1_APP}
    \mathcal{C}_2^*=\mathcal{C}_2(t\to \infty)= \frac{4D^2}{r^2}V(m,N), \qquad V(m,N)=\frac{N^2(m-1)}{m^2(2N-m-1)}.
\end{equation}

It is interesting to study the extrema of the function $V(m,N)$. To this end, we first treat $m$ as a continuous variable and analyze the stationary points obtained from the condition
\begin{equation}
\frac{dV(m,N)}{dm}=0 .
\end{equation}
The solutions of this equation provide candidate extrema in the continuous variable $m$. Since $m$ is restricted to integer values, the location of the maximum (or minimum) is then determined by evaluating $V(m,N)$ at the integer values closest to the stationary points, and by comparing these values with those at the boundaries $m=1$ and $m=N$. Note that the latter satisfy $V(1,N)=0$ and $V(N,N)=1$ for all $N\geq2$. Setting the derivative $\frac{dV(m,N)}{dm}$ equal to zero yields
\begin{equation}\label{eq:tofindmaxV}
N^2\, m\,[2m^2-2m(N+1)+4N-2]=0 ,
\end{equation}
which must be analyzed for $N\geq2$ and $1\leq m\leq N$. Discarding the trivial solution $m=0$, we are left with the quadratic equation
\begin{equation}\label{eq:tofindmaxVsecondeq}
m^2-m(N+1)+2N-1=0 .
\end{equation}
For $N>5$, \eqref{eq:tofindmaxVsecondeq} admits two real solutions
\begin{equation}
m_{\pm}=\frac{(N+1)\pm\sqrt{(N-1)(N-5)}}{2}.
\end{equation}
The larger root $m_+$ corresponds to a local minimum, while the smaller root $m_-$ corresponds to a maximum. Moreover, for $N\geq6$ one finds $2<m_-<3$. Since $m$ must be an integer, the maximum of $V(m,N)$ for $N\geq6$ must occur at
either $m=2$ or $m=3$. We therefore compare
\begin{equation}
V(2,N)=\frac{N^2}{4(2N-3)}, \qquad
V(3,N)=\frac{2N^2}{9(2N-4)}.
\end{equation}
A direct comparison shows that $V(2,N)>V(3,N)>1$ for all $N>6$. Hence, for $N>6$, the global maximum of $V(m,N)$ over the integers $m\in[1,N]$ is
attained at $m=2$.

For $N=6$, instead, one finds $V(2,6)=V(3,6)=V(6,6)=1$,
so that the maximum is not unique and it is attained for three distinct values of $m$.

For $N=5$, the two solutions $m_+$ and $m_-$ coincide and correspond to an inflection point. Thus, in this case, the function $V(m,5)$ is monotonically increasing in the interval $m\in[1,5]$, and its maximum is reached at the boundary $m=5$.

Finally, for $N<5$, \eqref{eq:tofindmaxVsecondeq} has no real solutions, indicating the absence of local extrema in the open interval $m\in(1,N)$. The function $V(m,N)$ is therefore monotonic, and its maximum is again attained at the boundary $m=N$.
\begin{figure}[h]
    \centering
    \includegraphics[width=0.6\textwidth]{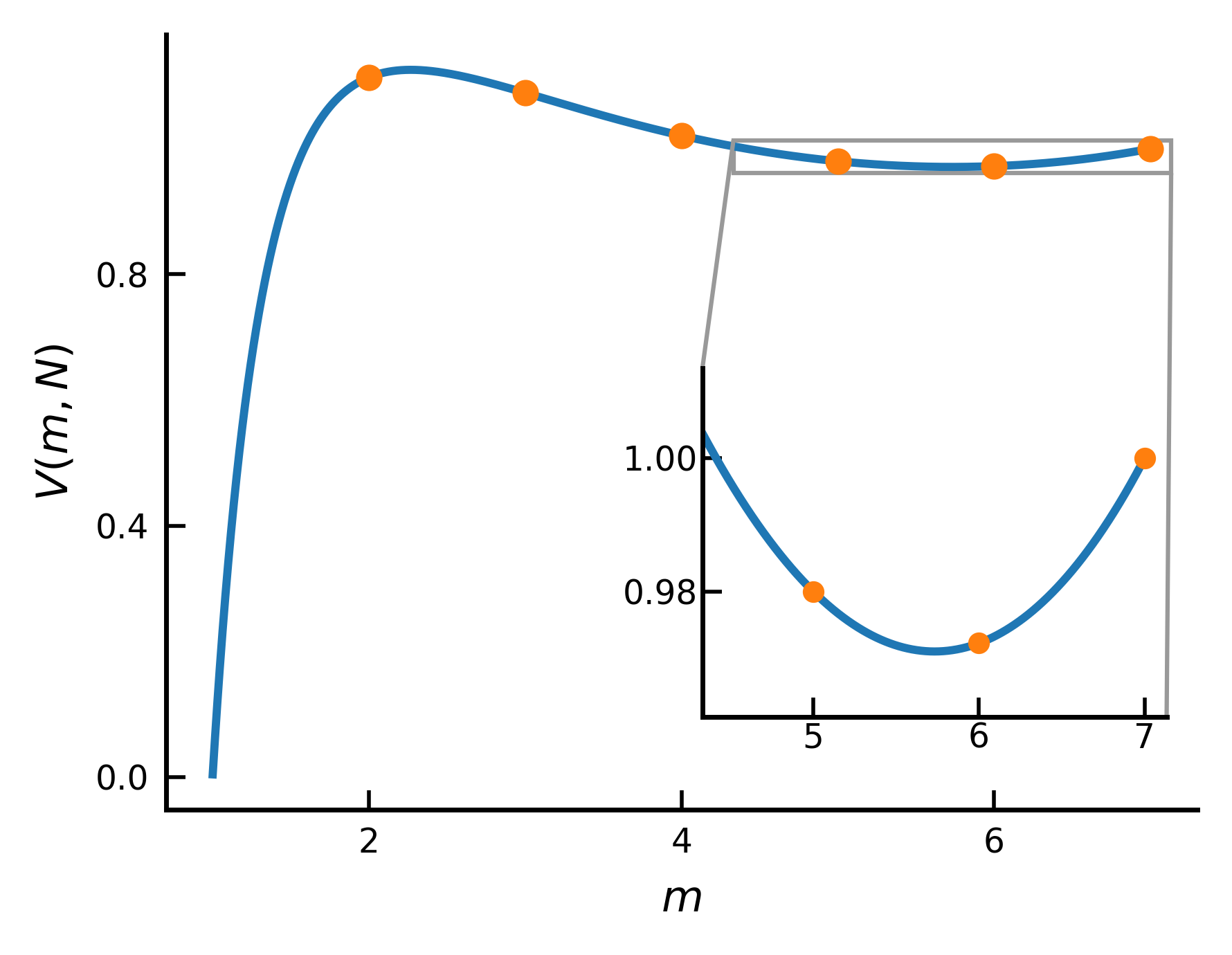} 
    \caption{Plot of the function $V(m,N=7)$ given in \eqref{eq:V(m,N)_d1_APP}. The inset clearly shows the local minimum at $m=N-1=6$. The orange dots are just for visual clarity and not the results of simulations.}
    \label{fig:mindiV(m,N)}
\end{figure}

As anticipated, the point \(m_+\) corresponds to a local minimum of the function \(V(m,N)\) when $N>5$.
Moreover, for \(N>5\) one has \(N-2<m_+<N-1\). Since \(m\) must take integer values, the discrete minimum in the vicinity of \(m_+\) must therefore occur either at \(m=N-2\) or \(m=N-1\). Evaluating \(V(m,N)\) at these two values yields
\begin{equation}
V(N-1,N)=\frac{N(N-2)}{(N-1)^2}, \qquad
V(N-2,N)=\frac{N^2(N-3)}{(N-2)^2(N+1)}.
\end{equation}
A direct comparison shows that $V(N-1,N)<V(N-2,N)<1,$ which implies that, for all \(N>5\), the function \(V(m,N)\) attains a local minimum at \(m=N-1\).
It is also useful to rewrite
\begin{equation}
V(N-1,N)=1-\frac{1}{(N-1)^2},
\end{equation}
which makes it explicit that this minimum lies strictly below unity, while its deviation from $1$ is small and it vanishes as $N\to\infty$.

This analysis shows that the function $V(m,N)$, which quantifies the strength of
stationary correlations, undergoes a qualitative change when the number of particles $N$ crosses the critical value $N_c=6$. For $N<N_c$, $V(m,N)$ increases
monotonically with $m$, whereas for $N>N_c$ it becomes non–monotonic. As a result, the global maximum shifts discontinuously from the boundary $m=N$ to the interior value $m=2$ as $N$ crosses $N_c$. In this regime, $V(m,N)$ also develops a shallow local minimum at $m=N-1$, whose
depth decreases with increasing $N$ and vanishes in the limit $N\to\infty$. At the critical point $N=N_c$, the maximum is degenerate, and
$V(m,N_c)$ attains the same maximal value $1$ for $m=2$, $m=3$, and $m=N$.

It is also interesting to study the large $N$ asymptotics of the function $V(m,N)$. It displays two distinct regimes, depending on how $m$ scales with $N$: 

\noindent\textbf{(i) $m=\kappa N$ with $0<\kappa\le 1$.}
If a finite fraction $\kappa$ of the particles is reset at each event, then
\begin{equation}
    V(m,N)\;\xrightarrow[N\to\infty]{}\;
\tilde v_1(\kappa)=\frac{1}{\kappa(2-\kappa)}.
\end{equation}
This limit is regular for any $0<\kappa<1$, while it becomes singular as $\kappa\to 0$: 
\begin{equation}
    \tilde v_1(\kappa)\sim\frac{1}{2\kappa}.
\end{equation}
This limit is noteworthy because $\tilde v_1(\kappa)$ diverges as $\kappa = m/N \to 0$, while for the extreme case $m=1$ (corresponding to $\kappa\equiv 0$) the function $V$ remains identically zero, i.e. $V(m=1,N)=0$, for all $N$.

\noindent\textbf{(ii) $m=O(1)$ as $N\to\infty$.}
When the number of reset particles does not grow with $N$, the behavior changes qualitatively. In this regime,
\begin{equation}
    V(m,N)\;\xrightarrow[N\to\infty]{}\; N\, \tilde v_2(m),
\qquad 
\tilde v_2(m)=\frac{m-1}{2m^2},
\end{equation}
which means that $V(m,N)$ grows linearly with the system size. Here it is also clear that the function $\tilde v_2(m)$ reaches a maximum at $m=2$.
For large $m$,
\begin{equation}
    \tilde v_2(m)\sim\frac{1}{2m}\qquad (m\to\infty),
\end{equation}
which, as expected, matches the small-$\kappa$ limit of $\tilde{v}_1(\kappa)$.

\section{Physical interpretation of the maximum of $A(z;\beta)$}
\label{sec:PhysicalInterpretationofthemaximum}

In order to understand, from a physical perspective, why the function $a(z=\tilde{r}t;m,N)=A(z;\beta)$ has a maximum, we can write a differential equation governing the time evolution of the quantities: $v(t)=\langle x_i^2(t) \rangle$, $\mathcal{C}_1(t)$, $\mathcal{C}_2(t)$ and $a(t;m,N)$. To do so, we can multiply \eqref{eq:FPeq_xspace_APP} by $x_i^2$ or $x_i^4$ or $x_i^2 x_j^2$, and integrate over all possible configurations $\mathbf{x}$.
The results read as follows:
\begin{align}
\frac{dv(t)}{dt} &= 2D - \tilde r\,v(t),
\qquad \tilde r = \frac{m}{N}\,r,
\label{eq:v_dot} \\[6pt]
\frac{d \mathcal{C}_1(t)}{dt} &= -\,\tilde r\,\mathcal{C}_1(t)
\;+\; 8D\,v(t) \;+\; \tilde r\,v(t)^2,
\label{eq:C1_dot} \\[6pt]
\frac{d\mathcal C_2(t)}{dt} &= -\,\tilde{r} \beta\,\mathcal C_2(t)
\;+\; (2-\beta)\,\tilde r\,v^2(t),
\qquad 
\beta=\frac{2N-m-1}{N-1},
\label{eq:C2_dot}
\end{align}
These equations can be used to derive a differential equation for \(a(t;m,N)=\mathcal C_2(t)/\mathcal C_1(t)\). For notational simplicity, in the following we write $a(t)$ instead of $a(t;m,N)$ whenever no confusion arises.
Using Eqs.~\eqref{eq:C1_dot} and \eqref{eq:C2_dot}, we obtain
\begin{equation}\label{eq:diff_eq_a}
    \frac{da(t)}{dt}
    =\frac{d}{dt}\frac{\mathcal C_2(t)}{\mathcal C_1(t)}
    =\frac{1}{\mathcal C_1(t)}\left[
    \frac{d\mathcal C_2(t)}{dt}
    - a(t)\,\frac{d\mathcal C_1(t)}{dt}
    \right]=-\,(\beta-1)\tilde r\,a(t)
    +(2-\beta)\tilde r\,\frac{v^2(t)}{\mathcal C_1(t)}
    -a(t)\,\frac{8D\,v(t)+\tilde r\,v^2(t)}{\mathcal C_1(t)} .
\end{equation}
We now need to solve \eqref{eq:v_dot} and \eqref{eq:C1_dot} explicitly.
The results read
\begin{equation}
    v(t)=\frac{2D}{\tilde r}\bigl(1-e^{-\tilde r t}\bigr),
    \quad \text{and} \quad
    \mathcal C_1(t)=\left(\frac{2D}{\tilde r}\right)^2
    f_1(\tilde r t),
    \quad
    f_1(z)=5-4e^{-z}-6z\,e^{-z}-e^{-2z}.
\end{equation}
We can now insert these expressions into \eqref{eq:diff_eq_a}.
Introducing the rescaled time \(z=\tilde r t\), the differential equation
for the scaling function
\(A(z;\beta)\equiv a(\tilde r t;m,N)\) reads
\begin{equation}\label{eq:deffeq_a_inz}
    \frac{dA(z;\beta)}{dz}
    =-\,\Big[(\beta-1)+\omega(z)+\phi(z)\Big]\,A(z;\beta)
    +(2-\beta)\,\phi(z),
\end{equation}
where
\begin{equation}\label{eq:PSi_Phi}
\omega(z)=\frac{4(1-e^{-z})}{f_1(z)},
\qquad
\phi(z)=\frac{(1-e^{-z})^2}{f_1(z)}.
\end{equation}
These functions behave as
\begin{equation}
    \phi(z)\approx 
    \begin{cases}
        \dfrac{1}{2}-\dfrac{z}{4},  & z\to0,\\[4pt]
        \dfrac{1}{5}+\dfrac{6}{25}(z-1)e^{-z}, & z\to \infty,
    \end{cases}
\end{equation}
\begin{equation}
    \omega(z)\approx
    \begin{cases}
        \dfrac{2}{z}+\dfrac{z}{6}, & z\to0,\\[4pt]
        \dfrac{4}{5}+\Big(\dfrac{24z-4}{25}\Big)e^{-z}, & z\to \infty.
    \end{cases}
\end{equation}

Equation~\eqref{eq:deffeq_a_inz} shows that the evolution of $A(z;\beta)$ is governed by two distinct types of contributions: negative terms proportional to $A(z;\beta)$, which erase pre-existing correlations, and a source term, $(2-\beta)\,\phi(z)$, which generates new correlations. The balance between these competing effects determines the overall correlation strength.
Diffusion acts as a purely decorrelating mechanism, encoded in the term $\omega(z)A(z;\beta)$: during diffusive motion, particles evolve independently and any existing correlations are progressively washed out. Resetting also destroys correlations, because whenever a reset occurs the past trajectory of the reset particle is forgotten. This loss of memory is captured by the term $\phi(z)A(z;\beta)$, which is present even when all particles are reset simultaneously.
However, resetting does not only erase correlations: it can also create them. When two particles are reset simultaneously, they are brought to the same position and correlations are reinjected into the system. This effect appears through the source term $(2-\beta)\phi(z)$, showing that the same function $\phi(z)$ that controls the loss of correlations due to resetting also controls their reinjection, provided the reset is synchronous. The prefactor $2-\beta$ reflects the probability that the two particles are reset simultaneously: indeed, the corresponding rate $\tilde r(2-\beta)= r\,m(m-1)/[N(N-1)]$ is precisely the probability per unit time that two fixed particles are reset in the same event.
An additional decorrelation mechanism arises only when $m\neq N$ ($\beta \neq 1$). In this case, a reset event may affect one particle but not the other, thereby destroying their mutual correlation without any possibility of reinjection. This process is encoded in the term $(\beta-1)A(z;\beta)$, whose rate $\tilde r(\beta-1)=r\,m(N-m)/[N(N-1)]$ is precisely the probability that only one of the two particles is reset. The competition between correlation injection by synchronous resetting and correlation loss due to diffusion and asynchronous resetting leads to the non-monotonic behavior of $A(z;\beta)$. When $m=N$, asynchronous resetting is absent, the term $\beta-1$ vanishes, and no decorrelation phase is observed. Thus, the additional decorrelating phase observed in batch resetting for $m\neq N$ originates entirely from asynchronous resetting events.

\section{Connected pair correlation function}
\label{sec:connectedpaircorrel}

We now consider the large $N$ limit, where our system of $N$ particles turns into a strongly correlated gas in the continuum. To probe the correlations in such systems, we need to go beyond the correlations at the individual particle levels and instead study the density-density correlation functions of the gas. This can be done by coarse-graining the system by first defining the empirical density (normalised to unity)
\begin{eqnarray} \label{def_rho}
\hat \rho(x) = \frac{1}{N}\sum_{i=1}^N \delta(x-x_i) \;.
\end{eqnarray}
Its average is indeed the coarse-grained density profile given by
\begin{eqnarray} \label{def_rho_av}
\langle \hat\rho(x) \rangle= \frac{1}{N} \left\langle \sum_{i=1}^{N} \delta(x - x_i) \right\rangle \;,
\end{eqnarray}
where the average is taken over the JPDF. Note that $\langle \hat\rho(x) \rangle$ then coincides with the one-point marginal discussed in the main text and also
in Section \ref{sec:FP_derivation}, denoted by $P_m^1(x)$. The connected pair correlation function is then defined as
\begin{equation}
    G(x,y)=\langle \hat\rho(x) \hat\rho(y) \rangle-\langle \hat\rho(x) \rangle \langle \hat\rho(y) \rangle \;.
\end{equation}
The function $G(x,y)$ can also be re-expressed in terms of the one- and two-point marginals as
\begin{equation} \label{Gxy_SM}
    G(x,y)=P_m^2(x,y)-P_m^1(x)P_m^1(y)+\frac{1}{N}[P_m^1(x) \delta(x-y)-P_m^2(x,y)].
\end{equation}
It can then be computed explicitly by using the results from Eqs.~\eqref{eq:1-point_marg_APP} and \eqref{eq:2-point_marg_APP}.
In the large $N$ limit the last term in Eq. (\ref{Gxy_SM}) drops out, leading to 
\begin{equation} \label{Gxy_largeN}
    G(x,y)\approx P_m^2(x,y)-P_m^1(x)P_m^1(y) \;.
\end{equation}
This pair correlation function also satisfies the sum rule
\begin{eqnarray} \label{sum_rule}
\int G(x,y)\, dy = 0 \;.
\end{eqnarray}
This simply follows from the definition in Eq. (\ref{Gxy_largeN}). Indeed, when one integrates over $y$, the two-point marginal reduces to a one-point marginal and the integral of the one-point function is unity. The pair-correlation function $G(x,y)$ in Eq. (\ref{Gxy_largeN}) can 
be expressed in the scaling form
\begin{equation}\label{eq:G(x,y)_SM}
    G(x,y)=\frac{1}{\ell_m^2}\mathcal{G}\left( \frac{x}{\ell_m}, \frac{y}{\ell_m} \right) \quad, \quad {\rm where} \quad \ell_m^2 = \frac{D}{\tilde r}  \quad, \quad {\rm with} \quad \tilde r = \frac{m}{N}\,r \;.
\end{equation}
\begin{figure}[t] 
    \centering
    \includegraphics[width=0.6\textwidth]{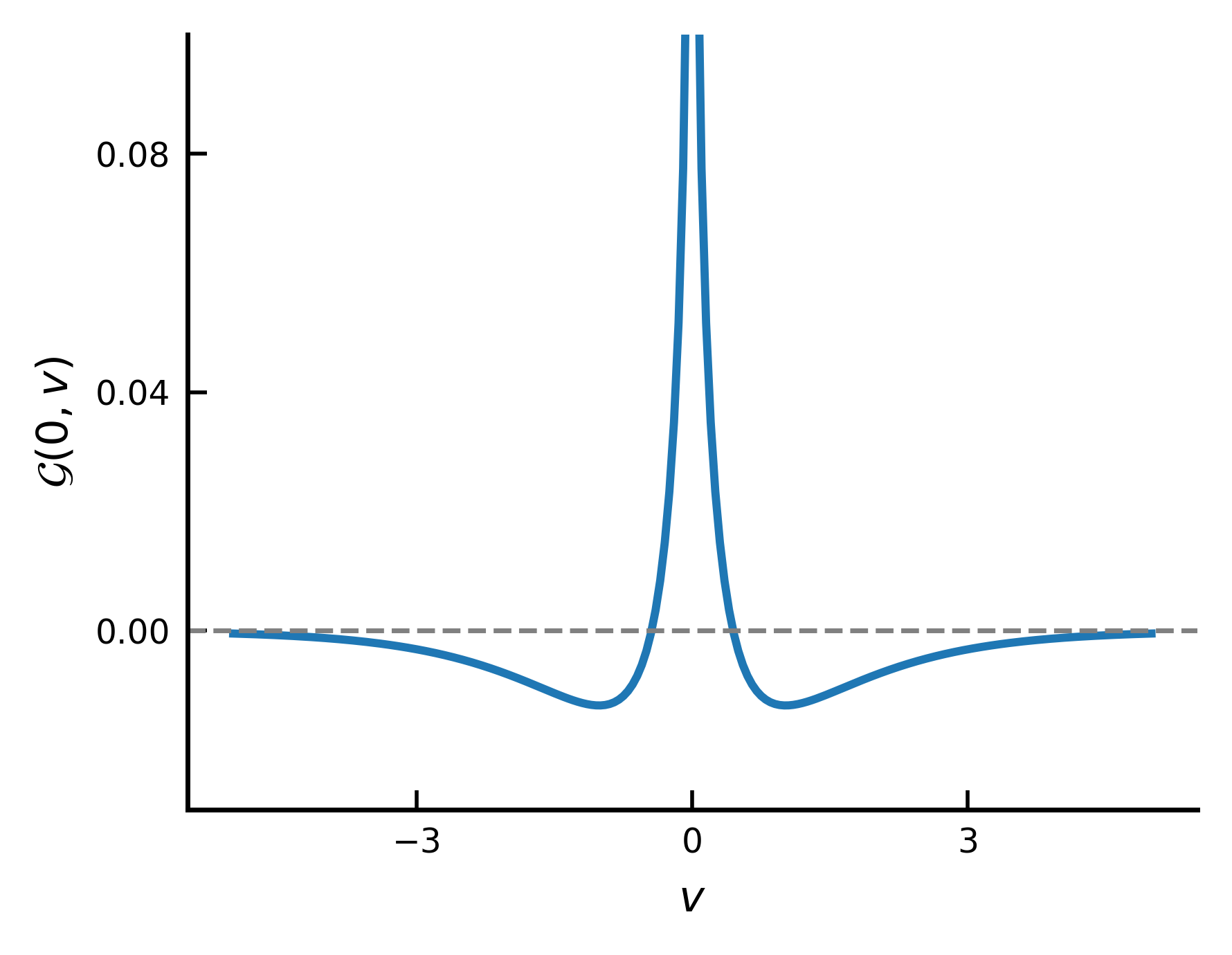} 
    \caption{Plot of $\mathcal{G}(u=0,v)$ [given in \eqref{Guv}, setting $u=0$] as a function of $v$.}
    \label{fig:G(0,y)}
\end{figure}
Using the results from Eqs.~\eqref{eq:1-point_marg_APP} and \eqref{eq:2-point_marg_APP} the scaling function is given by
\begin{equation} \label{Guv}
    \mathcal{G}(u,v)=\mathcal{P}_{\beta}(u,v)-\mathcal{R}(u) \mathcal{R}(v) \;,
\end{equation}
where $\mathcal{R}(u)$ and $\mathcal{P}_{\beta}(u,v)$ are defined respectively in \eqref{eq:1-point_marg_APP} and \eqref{eq:P(u,v)_APP}. Note that this function $ \mathcal{G}(u,v)$ depends on both coordinates $u$ and $v$ separately and not just on their difference, unlike in an isotropic and homogeneous gas. In Fig.~\ref{fig:G(0,y)}, we show a plot of the function $\mathcal{G}(0,v)$. This function is symmetric around $v=0$ and has the asymptotic behaviours  
\begin{equation}
\mathcal{G}(u=0,v)\approx
    \begin{cases}
    - \left(\frac{2-\beta}{2\pi}\ln \left(\frac{\sqrt{\beta}}{2}|v|\right) \right)\;, &v\to0 \;,\\[10pt]
    -\frac{1-\sqrt{\beta-1}}{4}e^{-v}, & v\to \infty \;.
    \end{cases}
\end{equation}
The logarithmic divergence at small $v$ is a direct signature of simultaneous resetting and, as expected, its amplitude is controlled by the factor $(2-\beta)$, which quantifies the probability of simultaneous reset events. 
Indeed, particles that are reset in the same event are reinjected at the same position, leading to an enhanced probability of finding particle pairs at vanishing separation and thus to a divergent short-distance behavior of the connected correlation function.
For $u=0$ and large $v$, the dominant contribution comes from asynchronous resetting events, in which one particle is reset to the origin ($u=0$) while the other evolves freely and moves far away, resulting in an exponential decay of correlations controlled by $(\beta-1)$.
Note that due to the sum rule in Eq. (\ref{sum_rule}), we have $\int_0^{\infty}dv \, \mathcal{G}(0,v)=0$, indicating that $\mathcal{G}(0,v)$ must change sign in some region of space, as seen in Fig. \ref{fig:G(0,y)}: thus the excess of probability at short distances induced by simultaneous resetting must be compensated by a depletion at larger separations.

\section{Spectral form factor}
\label{sec:SFF}
In this section, we study the spectral form factor of our system.
We have defined the empirical density of particles as
\begin{equation}
    \hat{\rho}(x)=\frac{1}{N} \sum_{i=1}^N \delta(x-x_i)
\end{equation}
Its FT $\tilde{\rho}(k)$ reads
\begin{equation}
    \tilde{\rho}(k)= \int_{-\infty}^{\infty}\! dx \, e^{ikx} \, \hat{\rho}(x)= \frac{1}{N} \sum_{i=1}^N e^{ikx_i}
\end{equation}
We can then define the spectral form factor ($SFF$) as
\begin{equation}
    SFF(k)= \frac{1}{N^2} \sum_{i,j} e^{ik(x_i-x_j)}= |\tilde{\rho}(k) |^2 \;.
\end{equation}
\begin{figure}[t] 
    \centering
    \includegraphics[width=0.6\textwidth]{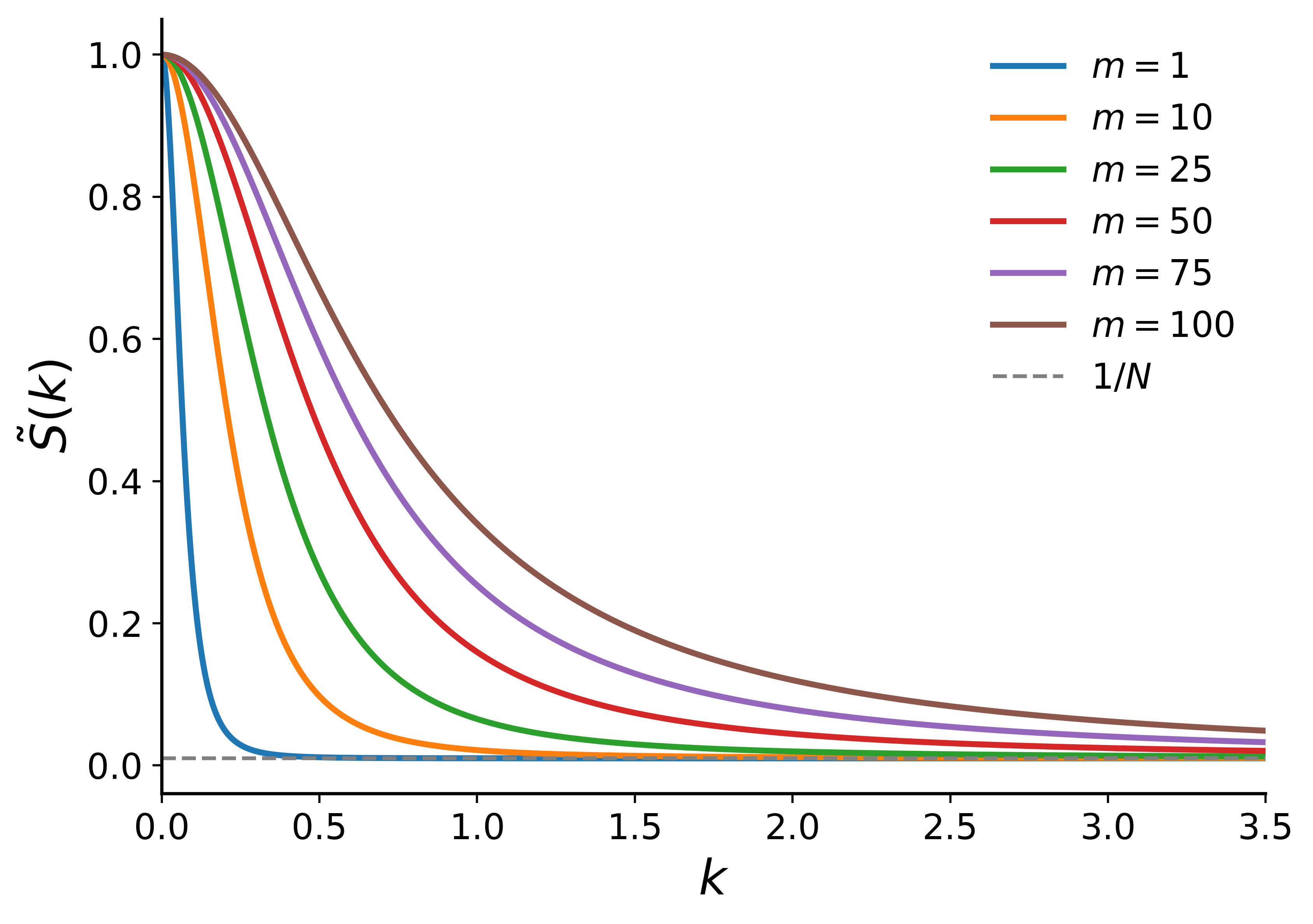}
    \caption{Plot of $\tilde{S}(k)$ [given in \eqref{eq:tildeS_SM}] for $N=100$ and different values of $m$.}
    \label{fig:SFF(k)}
\end{figure}
Since $SFF(k)$ depends on the random variable $x_i$, we consider its average 
\begin{equation}\label{eq:tildeS(k)_def}
    \tilde{S}(k) = \langle SFF(k)\rangle= \frac{1}{N^2} \sum_{i,j} \langle e^{ik(x_i-x_j)} \rangle
\end{equation}
This function captures the typical strength of the density fluctuations at wavelength $\frac{2\, \pi}{k}$.
Eq. \eqref{eq:tildeS(k)_def} can be rewritten in terms of the FT of the 2-point marginal defined in \eqref{eq:P2tilde_APP} as
\begin{equation}\label{eq:tildeS_SM}
    \tilde{S}(k)= \frac{N-1}{N} \tilde{P}_m^2(k,-k) + \frac{1}{N}
\end{equation}
 where
\begin{equation}
\tilde{P}_m^2(k_1,k_2) = \frac{\,(2-\beta) + (\beta-1) \left[\tilde{P}_m^1(k_1) + \tilde{P}_m^1(k_2)\right]}{ \,  \, \beta + \ell_m^2(k_1^2 + k_2^2)},\quad \text{with} \quad \tilde{P}_m^1(k) = \frac{1}{1+\ell_m^2 k^2},
\end{equation}
$\beta=\frac{2N-m-1}{N-1}$, $\ell_m= \sqrt{\frac{D}{\tilde{r}}}$ and $\tilde{r}= \frac{m}{N}r$.
This function behaves as
\begin{equation}
    \tilde{S}(k)\simeq
    \begin{cases}
        1-2\frac{N-1}{N}\,\ell_m^{2}\,k^{2}+O(k^{4}), & k\to0,\\
\frac{1}{N}
+\frac{1}{2}\frac{m-1}{N}\,
\frac{1}{\ell_m^2}\,
\frac{1}{k^{2}}
+O\!\left(\frac{1}{k^{4}}\right), & k\to \infty.
    \end{cases}
\end{equation}

To see the role played by inter-particle correlations, it is instructive to compare \eqref{eq:tildeS_SM} with the spectral form factor of a reference system in which the $N$ particles are independent and identically distributed according to the same one-point marginal \eqref{eq:1-point_marg_APP}. In this uncorrelated case, all nontrivial two-point information is lost, and the averaged spectral form factor reduces to
\begin{equation}
    \tilde{S}_{\rm iid}(k)= \frac{1}{N}+\frac{N-1}{N} \frac{1}{(1+\ell_m^2k^2)^2}.
\end{equation}
The asymptotic behavior of $\tilde{S}_{\rm iid}(k)$ is given by
\begin{equation}
    \tilde{S}_{\rm iid}(k)\simeq
    \begin{cases}
        1 - \frac{2(N-1)}{N}\,\ell_m^{2}\,k^{2}
    + O(k^{4}), & k\to0,\\
    \frac{1}{N}
    + \frac{N-1}{N}\,\frac{1}{\ell_m^{4}\,k^{4}}
    + O(k^{-6}), & k\to \infty.
    \end{cases}
\end{equation}
Comparing the two cases, we observe that $\tilde S(k)$ and $\tilde S_{\mathrm{iid}}(k)$ share the same small-$k$ expansion, as expected since they are constructed from identical one-point marginals. However, their large-$k$ behavior is qualitatively different. In particular, the slower $k^{-2}$ decay of $\tilde S(k)$, as opposed to the $k^{-4}$ decay of $\tilde S_{\mathrm{iid}}(k)$, provides a clear spectral signature of the correlations induced by batch resetting.

\section{Generalization to higher dimensions $d$}

In this section we extend our analysis to Brownian particles evolving in $d$ spatial dimensions. We consider a system of \(N\) such particles, where the position of each one is described by a vector \(\mathbf{x}_i=(x_{i,1},\ldots,x_{i,d}) \in \mathbb{R}^d\), where $i=1,...,N$ labels the particle. We also define the vector $\mathbf{X}=(\mathbf{x_1},...,\mathbf{x}_N)\in \mathbb{R}^{dN}$.
In this setting, the Fokker--Planck equation in \eqref{eq:FPeq_xspace_APP} generalizes to
\begin{equation}\label{eq:FP_xspace_d_dim}
\begin{aligned}
\partial_t P_m(\mathbf X,t)
= D \sum_{i=1}^{N} \nabla_{\mathbf x_i}^2
P_m(\mathbf X,t)- r\, P_m(\mathbf X,t)+ \frac{r}{\binom{N}{m}}
\sum_{\substack{S_m}}\left(\prod_{i \in S_m} \delta^{(d)}(\mathbf x_i)\right)\,P_m^{(N-m)}(\mathbf X_{S_m^c},t),
\end{aligned}
\end{equation}
Here we have defined $\nabla_{\mathbf{x}_i}^2
= \sum_{\alpha=1}^{d} \frac{\partial^2}{\partial x_{i,\alpha}^2}$ and $\delta^{(d)}(\mathbf{x}_i)
= \prod_{\alpha=1}^{d} \delta(x_{i,\alpha})$. As in the $d=1$ case studied above, the set $S_m$ is a subset 
$S_m\subseteq\{1,\ldots,N\}$ of size $|S_m|=m$, which is selected uniformly at random at each reset event. The set $S_m^c$ is the complementary set, i.e. $S_m^{\mathrm{c}} = \{1,\ldots,N\} \setminus S_m$. The vector $\mathbf X_{S_m^c}$ is obtained from the vector $\mathbf{X}$ by marginalizing over the vectors $\mathbf{x}_i$ for all $i\in S_m$.
We have also defined the JPDF $P_m(\mathbf x_1,\ldots,\mathbf x_N,t)=P_m(\mathbf X,t)$, which now is a function of $N d +1 $ variables. The function $P_m^{(N-m)}$ indicates the $(N-m)$-particles marginal, i.e. the JPDF marginalized over all the $d$ components of $m$ different particles, namely
\begin{equation}
    P_m^{(N-m)}(\mathbf{X}_{S_m^c},t)=\!\int\!\!\int d\mathbf{X}_{S_m}P_m(\mathbf{X},t),
\end{equation}
where $d\mathbf{X}_{S_m}=\prod_{i\in S_m} d\mathbf{x}_i=\prod_{i\in S_m} \prod_{j=1}^d dx_{i,j}$. 
Taking the FT of \eqref{eq:FP_xspace_d_dim}, defined as
\begin{equation}
\tilde P_m(\mathbf k_1,\ldots,\mathbf k_N,t)
=\int \!\! \int \prod_{i=1}^{N} d^d \mathbf x_i\;
\exp\!\left(i \sum_{i=1}^{N} \mathbf k_i \cdot \mathbf x_i
\right)P_m(\mathbf x_1,\ldots,\mathbf x_N,t),
\end{equation}
we obtain
\begin{equation}\label{eq:FP_Kspace_d_dim}
\partial_t \tilde P_m(\mathbf K,t)
= -\left( D \sum_{i=1}^{N} |\mathbf k_i|^2 + r \right)
\tilde P_m(\mathbf K,t)
+ \frac{r}{\binom{N}{m}}
\sum_{\substack{S_m}}
\tilde P^{N-m}_m(\mathbf K_{S_m^c},t),
\end{equation}
where we have set $\mathbf{K}=(\mathbf{k}_1,...,\mathbf{k}_N)\in \mathbb{R}^{Nd}$ and $\tilde P_m(\mathbf k_1,...,\mathbf{k}_N,t)=\tilde P_m(\mathbf K,t)$. The vector $\mathbf K_{S_m^c}$ is obtained from $\mathbf{K}$ by setting the vector $\mathbf{k}_i=\mathbf{0}$ for all $i\in S_m$ and $\tilde P^{N-m}_m$ is the FT of the $(N-m)$-particles marginal.

\subsection{Correlations}
The correlators introduced in the Letter and defined in
\eqref{eq:defdiCij} extend straightforwardly to particles evolving
in $d$ spatial dimensions. We define
\begin{equation}
    \mathcal C_{1,d}(t)
    =
    \langle |\mathbf x_i|^4(t) \rangle
    -
    \langle |\mathbf x_i|^2(t) \rangle^2,
\end{equation}
while, for $i\neq j$,
\begin{equation}
    \mathcal C_{2,d}(t)
    =
    \langle |\mathbf x_i|^2(t)\,|\mathbf x_j|^2(t) \rangle
    -
    \langle |\mathbf x_i|^2(t) \rangle
    \langle |\mathbf x_j|^2(t) \rangle.
\end{equation}
The normalized correlator then reads
\begin{equation}
    a_d(t;m,N)=\frac{\mathcal C_{2,d}(t)}{\mathcal C_{1,d}(t)}.
\end{equation}
\begin{figure}[t]
\centering
\begin{minipage}[t]{0.33\linewidth}
  \centering
  \includegraphics[width=\linewidth]{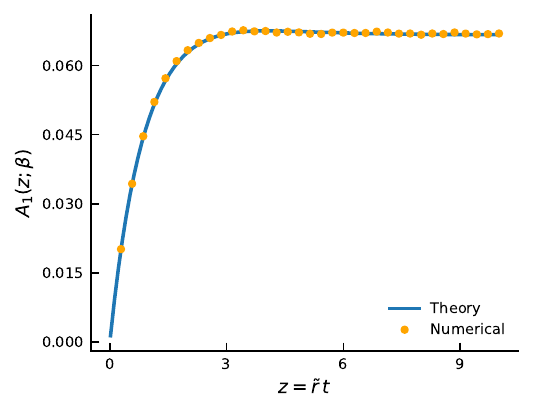}\\[-2mm]
  {\small (a) $d=1$}
\end{minipage}\hfill%
\begin{minipage}[t]{0.33\linewidth}
  \centering
  \includegraphics[width=\linewidth]{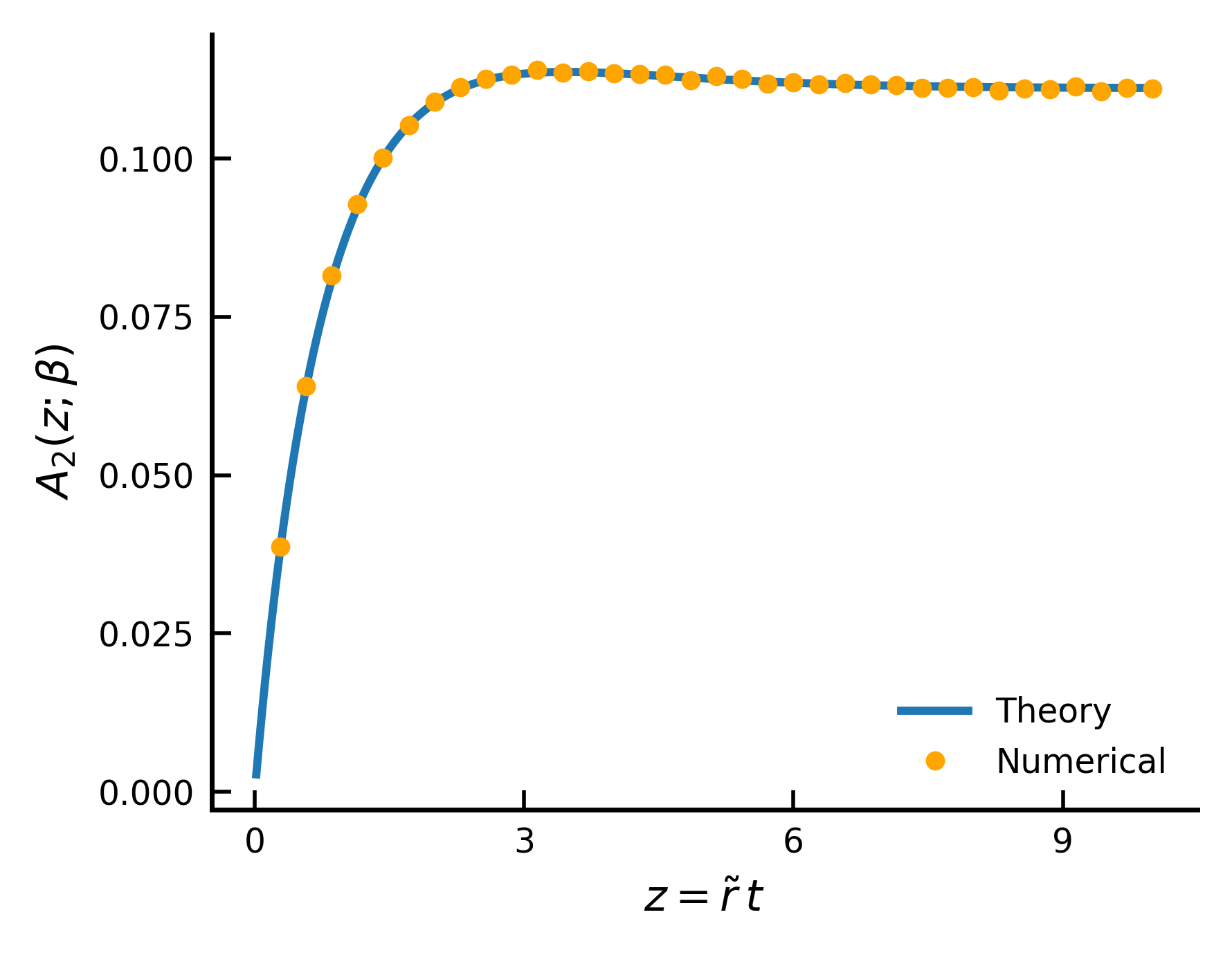}\\[-2mm]
  {\small (b) $d=2$}
\end{minipage}\hfill%
\begin{minipage}[t]{0.33\linewidth}
  \centering
  \includegraphics[width=\linewidth]{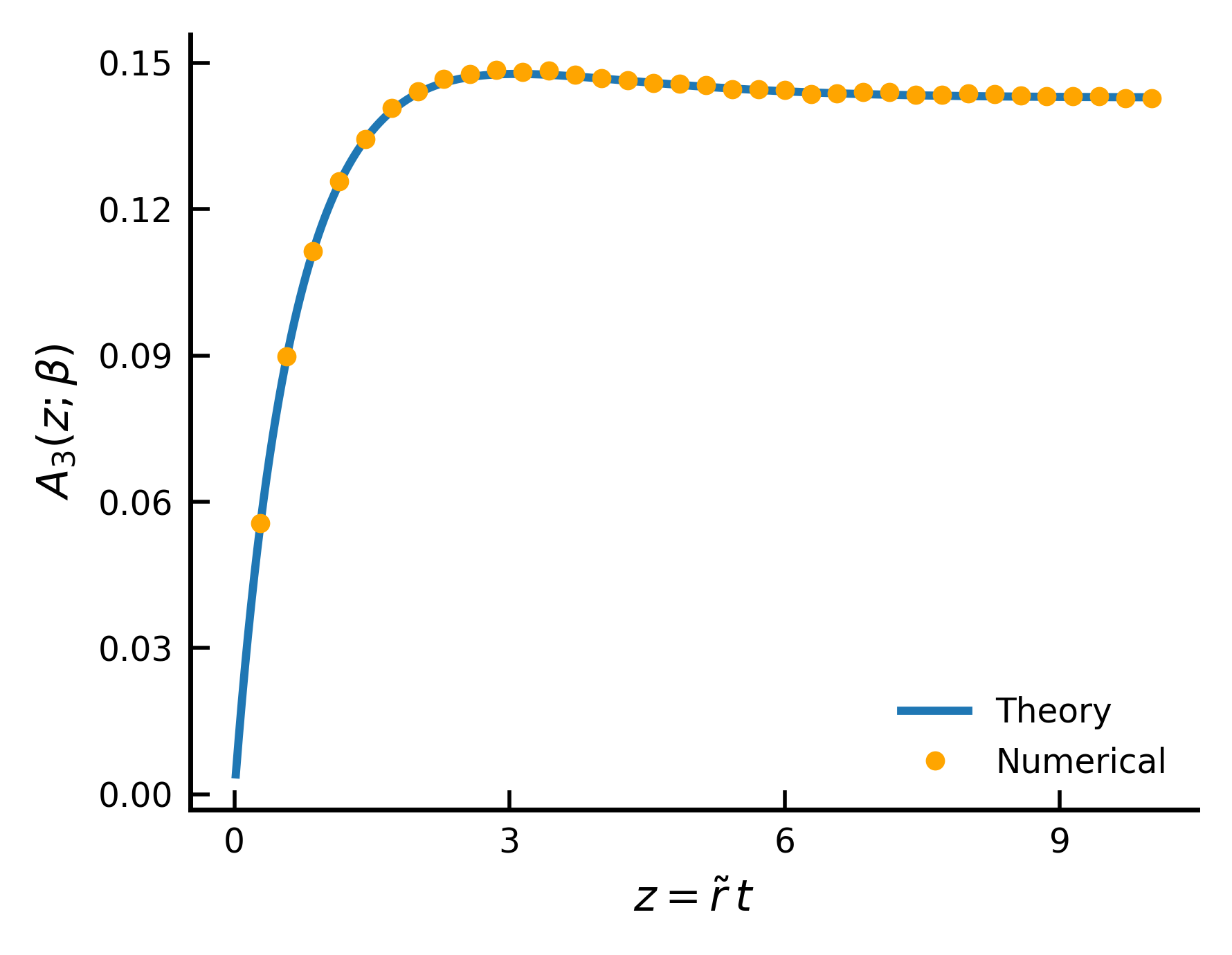}\\[-2mm]
  {\small (c) $d=3$}
\end{minipage}
\caption{Comparison between the results of numerical simulations (orange dots) and the theoretical predictions for the function $A_d(z;\beta)$ given in \eqref{eq:A_d_completeform}. The three panels correspond respectively to $d=1$, $d=2$ and $d=3$ for $\beta = 1.5$.}
\label{fig:Ad_d2_d3}
\end{figure}
Repeating the calculation of Sec.~\ref{sec:correlators}, starting from the Fokker--Planck equations
\eqref{eq:FP_xspace_d_dim} or \eqref{eq:FP_Kspace_d_dim}, we obtain
\begin{equation}\label{eq:C1_generald}
    \mathcal C_{1,d}(t)=\frac{4D^2}{\tilde{r}^2}\, f_{1,d}(\tilde r t), \quad \text{where} \quad  f_{1,d}(z)= d(d+4)-2d(2+z(2+d))\,e^{-z}-d^2 e^{-2z},
\end{equation}
and
\begin{equation}\label{eq:C2_generald}
    \mathcal C_{2,d}(t)=  \frac{4D^2}{\tilde{r}^2}
    f_{2,d}(\tilde{r} t;\beta), \quad \text{where} \quad f_{2,d}(z;\beta)=d^2\left[
    2\!\left(
    \frac{1 - e^{-\beta z}}{\beta}
    -\frac{e^{-z} - e^{-\beta z}}{\beta - 1}
    \right)
    - 1 - e^{-2z} + 2 e^{-z}
    \right].
\end{equation}
We recall that $z = \tilde{r}t$ with $\tilde{r} = \frac{m}{N} r$ and $\beta=\frac{2N-m-1}{N-1}$. We clearly see from \eqref{eq:C2_generald} that the function $\mathcal C_{2,d}(t)$ scales trivially with $d$, i.e. $\mathcal C_{2,d}(t)= d^2 \mathcal C_{2,1}(t)$, while $\mathcal C_{1,d}(t)$ acquires a nontrivial $d$-dependence.
By setting $d=1$ in \eqref{eq:C1_generald} and \eqref{eq:C2_generald} we recover the results in \eqref{eq:C1_APP} and \eqref{eq:C2_APP}.

Taking the long-time limit of Eq.~\eqref{eq:C2_generald}, we obtain the stationary correlator
$\mathcal{C}_{2,d}^*=\mathcal{C}_{2,d}(t\to \infty)$, namely
\begin{equation}
    \mathcal{C}_{2,d}^*
    = d^2 \frac{4D^2}{r^2} V(m,N),
    \qquad
    V(m,N)=\frac{N^2(m-1)}{m^2(2N-m-1)}.
\end{equation}
The function $V(m,N)$ coincides with the one-dimensional result
[Eq.~\eqref{eq:V(m,N)_d1_APP}], as it is independent of $d$.
As a consequence, the critical particle number $N_c=6$ remains unchanged in all dimensions and all the remarks made for $d=1$ in Sec. \ref{sec:correlators} remain valid in general.

We can also obtain the scaling function
$A_d(z;\beta)=a_d(\tilde r t;m,N)$, which reads
\begin{equation}\label{eq:A_d_completeform}
    A_d(z;\beta)=\frac{f_{2,d}(z;\beta)}{f_{1,d}(z)}=
    \frac{
    d\!\left[
    2\!\left(
    \frac{1 - e^{-\beta z}}{\beta}
    -\frac{e^{-z} - e^{-\beta z}}{\beta - 1}
    \right)
    - 1 - e^{-2z} + 2 e^{-z}
    \right]}{
    (d+4)-2(2+z(2+d))\,e^{-z}-d e^{-2z}} \;.
\end{equation}
By setting $d=1$ in the above expression, we recover \eqref{eq:a_APP}. 
From this expression, we find that in any dimension $d$ the function $A_d(z;\beta)$ displays a maximum at a finite $z^*(\beta)$, for any $1<\beta<2$. In Fig. \ref{fig:Ad_d2_d3} we show \eqref{eq:A_d_completeform} for $d=2$ and $d=3$ together with the results of numerical simulations.

Repeating the same analysis done in Sec.~\ref{sec:correlators} for $d=1$, we can obtain the value of the maximum analytically in the two limits $\beta \to 1^+$ and $\beta \to 2^-$.\\
(i) When $\beta \to 1^+$, the location of the maximum $z^*(\beta)$ behaves as $z^*(\beta)\simeq\frac{a_0(d)}{\beta-1}+a_1$, where $a_0(d)$ and $a_1$ are the solutions of the system
\begin{align}\label{eq:systema0a1_gend}
&\left(4 - 2a_0 + d - a_0 d\right) e^{a_0} = 4+d \\
& \left(2a_0 + 2a_1 - 2a_0 a_1 + d + a_0 d - a_0 a_1 d\right) e^{a_0} = 4+d .
\end{align}
It turns out that $a_1=2$ exactly and for all dimensions $d$.
In contrast, $a_0=a_0(d)$ depends on $d$, and it is the solution of \eqref{eq:systema0a1_gend}.
Defining $c = \frac{d + 4}{d + 2} > 1 $,\eqref{eq:systema0a1_gend} can be rewritten as
\begin{equation}
    (c - a_0) e^{a_0} = c.
\end{equation}
We can now express $a_0(d)$ as
\begin{equation}
a_0 = c + W\!\left(-c e^{-c}\right),
\end{equation}
where $W(z)$ is the Lambert $W$ function,  defined as the solution of $z=W(z) e^{W(z)}$. In the large $d$ limit we can set $c=\frac{d+4}{d+2}= 1+\epsilon$ with $\epsilon=\frac{2}{d+2}$. 
We can then use the expansion (see \cite{Corless1996})
\begin{equation}
    W\left(- \frac{1}{e}+\delta \right)\approx-1+\sqrt{2e\delta}
\end{equation}
valid for small $\delta$, where  $\delta=\frac{\epsilon^2}{2e}$ in our case. This gives us $W\!\left(-c e^{-c}\right)\approx-1+\epsilon$ and so
\begin{equation}
    a_0(d)\simeq 2\epsilon \simeq \frac{4}{d}.
\end{equation}
in the limit of large $d$.\\
\begin{figure}[h!] 
    \centering
    \includegraphics[width=0.6\textwidth]{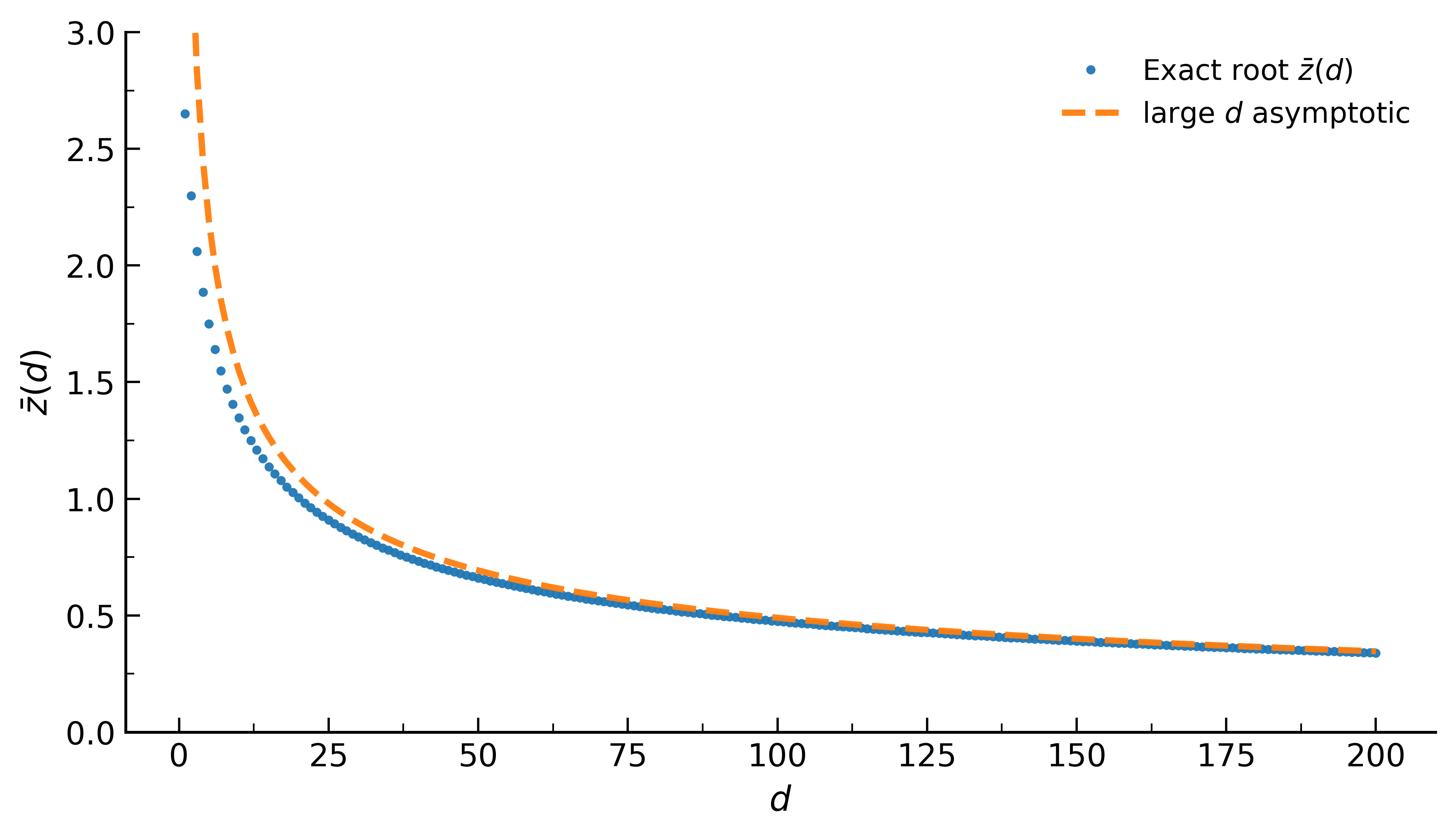}
    \caption{
    The blue dots represent the numerical results for $\bar z(d)$, obtained by solving \eqref{eq:zbartosolve}. The orange dashed line shows the large-$d$ asymptotic approximation of $\bar z(d)$, as given by \eqref{eq:larged_asympt_approx}.}
    \label{fig:barz_gend}
\end{figure}
(ii) In the opposite limit ($\beta\to 2^-$), we find that $z^*(\beta)$ converges to a finite value $z^*(\beta \to 2^-)=\bar z(d)$ for any finite $d$, where $\bar z(d)$ is the solution of the equation
\begin{equation}\label{eq:zbartosolve}
     e^{3z}(8 + 3d - (2 + d)z)
- e^{2z}(16 + 7d + 2(4 + d)z)
+ e^z\left(8 + 2z(5 + 2z) + d(5 + z(3 + 2z))\right)=d \;.
\end{equation}
In Fig. \ref{fig:barz_gend} we show the numerical solution of \eqref{eq:zbartosolve} as a function of $d$.
From the numerical analysis, we observe that the solution $\bar z(d)$ decreases as $d$ increases, suggesting that $\bar z(d)\to 0$ in the limit $d\to\infty$. We can therefore expand \eqref{eq:zbartosolve} for small $z$, obtaining
\begin{equation}\label{eq:expandedzbard}
\frac{2}{3} z^{4} + z^{5} + \left( \frac{34}{45} - \frac{d}{36} \right) z^{6}
+ O(z^{7})=0 \;.
\end{equation}
We assume a power-law scaling of the form $\bar z(d)\sim d^{-\alpha}$ for large $d$. Substituting this ansatz into \eqref{eq:expandedzbard}, the different contributions scale as
$z^{4}\sim d^{-4\alpha}$, $z^{5}\sim d^{-5\alpha}$, and $d z^{6}\sim d^{1-6\alpha}$.
A nontrivial solution requires a dominant balance between two leading terms. Balancing the $O(z^{4})$ and $O(d z^{6})$ contributions fixes the exponent to be $\alpha=\tfrac{1}{2}$.
With this choice, all remaining terms are subdominant in the limit $d\to\infty$, confirming the self-consistency of the scaling ansatz. We therefore set
\begin{equation}
\bar z(d)\simeq \frac{c_0}{\sqrt{d}}, \qquad d\gg 1.
\end{equation}
Substituting this expression back into \eqref{eq:expandedzbard} and retaining only the leading-order terms yields an algebraic equation for the prefactor $c_0$, whose nontrivial solution is $c_0=2\sqrt{6}$. We thus obtain the asymptotic behavior
\begin{equation}\label{eq:larged_asympt_approx}
\bar z(d)\simeq \frac{2\sqrt{6}}{\sqrt{d}}, \qquad d\to\infty.
\end{equation}
This asymptotic prediction, together with the numerical determination of $\bar z(d)$, is shown in Fig.~\ref{fig:barz_gend}.